\documentclass[trackchanges,twocolumn]{aastex701}
\usepackage{xcolor}
\usepackage{amsmath}

\begin{document}

\title{Odds Estimation for Supermassive Black Hole Binaries in Local Core Galaxies}

\author[orcid=0000-0002-4231-7802,sname='CJ']{C.J. Harris}
\affiliation{University of Michigan Department of Astronomy \& Astrophysics}
\email[show]{cordellh@umich.edu}  

\author[orcid=0000-0002-1146-0198, sname='Kayhan']{Kayhan Gültekin} 
\affiliation{University of Michigan Department of Astronomy \& Astrophysics}
\email{kayhan@umich.edu}

% \author[orcid=0000-0002-2183-1087, sname='Laura']{Laura Blecha} 
% \affiliation{University of Florida, Department of Physics}
% \email{kayhan@umich.edu}

% \collaboration{all}{The Terra Mater collaboration}

\begin{abstract}
Supermassive black hole binaries (SMBHBs) are expected products of massive galaxy mergers, but identifying binaries in nearby galaxies remains challenging. We combine simulated massive-galaxy assembly histories from IllustrisTNG with a semi-analytic core-scouring model to connect observable core galaxy properties to the probability of hosting a bound SMBHB at z=0. We (i) construct a core galaxy population using an observationally motivated galaxy stellar mass function, (ii) model binary evolution through dynamical friction, stellar hardening, and gravitational-wave emission, and (iii) perform logistic regressions with five observable predictors: stellar mass deficit, stellar mass, central density, central velocity dispersion, and effective radius. The resulting models perform well on simulated systems, achieving a pseudo-$R^2$ of $0.51\pm0.03$, accuracy $0.95\pm0.01$, precision $0.92\pm0.03$, and recall $0.82\pm0.04$. Uncoalesced binaries are typically compact, with separations peaking near 0.3 pc, though only $\sim0.3\%$ emit in the pulsar timing array band. Applying the model to observed core galaxies, we identify eight nearby galaxies more likely than not to host SMBHBs. Of those eight host candidates, we note Messier 86 (NGC 4406) as a particularly likely example that was similarly identified using independent means and therefore warrants future follow-up.
\end{abstract}



\section{Introduction} 

Supermassive black holes (SMBHs) are understood to be located in the centers of nearly all massive galaxies \citep{1998AJ....115.2285M,1998Natur.395A..14R,2005SSRv..116..523F}, with evidence in the form of direct imaging of horizon-scale emission \citep{2019ApJ...875L...1E}, observations of active galactic nuclei, and empirical scaling relations between black hole mass and host galaxy properties \citep{2007ApJ...662..808L,2009ApJ...698..198G,2013ARA&A..51..511K,2013ApJ...764..184M}. At the same time, massive galaxies grow hierarchically through mergers in $\Lambda$CDM cosmology. Since both progenitors may host central SMBHs, it is natural to ask what happens to those black holes after the galactic nuclei have merged.

The standard picture of SMBH pair evolution proceeds through several stages \citep{1980Natur.287..307B}. At large (kiloparsec) separations the black holes sink toward the center of the remnant merger primarily through dynamical friction against the background stars, gas, and dark matter. In this regime the \citet{1943ApJ....97..255C} prescription provides a useful description of the orbital decay. Below separations of order $\sim 100$ pc the two black holes become gravitationally bound, forming a binary, and dynamical friction ceases to be an efficient driver of evolution. Once a hard binary has formed, further orbital decay is caused by interactions with stars on centrophilic orbits \citep{1976MNRAS.176..633F,1996NewA....1...35Q,2006ApJ...651..392S,2013CQGra..30x4005M}. Stars that pass sufficiently close to the binary extract orbital energy and angular momentum through three-body encounters and are ejected from the nucleus. This process is referred to as loss-cone scattering, stellar scattering, or stellar hardening interchangeably.

The efficiency of stellar scattering depends on the rate at which the loss-cone is refilled \citep{1999MNRAS.309..447M,2017MNRAS.464.3131K}. In a spherical, collisionless galaxy loss-cone depletion can slow binary evolution and lead to the final parsec problem \citep{2003AIPC..686..201M, 2006ApJ...648..976M, 2013CQGra..30x4005M}. However, departures from spherical symmetry, including axisymmetry and triaxiality, can efficiently replenish loss-cone orbits \citep{2006astro.ph..1520H,2012ApJ...749..147K, 2015MNRAS.454L..66S, 2015ApJ...810...49V,2017MNRAS.464.2301G, 2021ApJ...922...40D}.

In gas-rich remnants, circumbinary disk (CBD) torques may also contribute to binary evolution \citep{2011MNRAS.411.1467K, 2013MNRAS.436.2997D, 2023MNRAS.522.2707S}. Many hydrodynamical simulations of accreting binaries indicate that material will preferentially accrete onto the secondary, enhancing the mass ratio while extracting angular momentum from the binary \citep{2020ApJ...901...25D, 2020MNRAS.498..537S, 2025ApJ...994..168C}.

Finally, at subparsec separations the binary becomes largely detached from the environment and dissipates energy purely through gravitational-wave emission \citep{1964PhRv..136.1224P}. Orbital decay at this stage is extremely efficient and drives rapid coalescence.

Despite significant theoretical progress, direct observation of SMBH binaries (SMBHBs) remains difficult. There is one confirmed parsec-scale binary in the literature \citep{2006ApJ...646...49R}, though many candidates exist \citep{2012ApJS..201...23E,2015ApJS..221....7R,2016MNRAS.463.2145C,2017MNRAS.468.1683R,2019NewAR..8601525D,2025arXiv250506221M,2025arXiv250816534A, 2026AAS...24725805M}. There are many more confirmed and candidate dual AGN \citep{2012AJ....143..119I, 2016MNRAS.456.1595M,2019MNRAS.483.4242L, 2019NewAR..8601525D,2019ApJ...877...17F,2020ApJ...892...29F,2021ApJ...907...71F,2020ApJ...899..154S,2021ApJ...922...83T,2025ApJ...993..151B, 2026arXiv260922433U}, but the connection between these observed systems and the underlying population of hard binaries remains uncertain, beyond the fact that dual AGN are in some sense a precursor to SMBHBs. The recent evidence for a nanohertz gravitational wave background (GWB) from pulsar timing array experiments \citep{2023ApJ...951L...6R, 2023A&A...678A..50E,2023ApJ...951L...8A,2023RAA....23g5024X} has provided important new constraints on the cosmic population of SMBHBs. Population models based on the inferred GWB indicate that binaries are more massive, more numerous, or more efficiently hardened than expected \citep{2023ApJ...952L..37A, 2024PhRvD.110f3020S, 2024ApJ...971L..29L}. In particular, the hardening timescale and its underlying physical mechanisms remain the least constrained components in predictions for the GWB \citep{2025PhRvD.111f3067L}.

Local core galaxies, massive gas-poor ellipticals exhibiting shallow inner surface-brightness profiles relative to the inward extrapolation of the outer envelope \citep{1997AJ....114.1771F,2006ApJS..164..334F,2007ApJ...664..226L,2009ApJS..181..486H}, offer a promising path toward linking electromagnetic observations with the SMBHB population. These cores are widely interpreted as the consequence of black hole binary scouring on both observational and theoretical grounds. Observationally, core galaxies exhibit tight scaling relations between the size of the core and the mass and influence radius of the central SMBH \citep{2007ApJ...662..808L, 2016Natur.532..340T, 2025arXiv251204178D}. The kinematic structure of core galaxies in general displays a coherent lack of stars on radial orbits in the core region and a uniform excess of radial orbits outside of it, in quantitative agreement with SMBHB simulations \citep{2014ApJ...782...39T}. Further support comes from N-body experiments showing the mass ejected from the core scales as the mass of the binary and number of dry mergers \citep{2006ApJ...648..976M, 2012ApJ...749..147K,2021ApJ...922...40D}. Previous studies have leveraged these ideas to constrain properties of the binaries responsible for forming cores as well as their hardening rates \citep{2024MNRAS.528....1H, 2026arXiv260107762H}, directly linking observed stellar cores to binary evolution.

Massive early-type galaxies (ETGs) are also exceptional candidates for hosting SMBHBs in the present epoch. The occupation fraction of parsec-scale SMBHBs may be as high as 50\% in $M_\star>10^{11}M_\odot$ galaxies, dropping to less than 10\% for less massive systems, with 75\% of SMBHBs of mass $M_{\rm tot}>10^9 M_\odot$ residing in classical bulges \citep{2023MNRAS.519.2083I}. Around 85\% of nanohertz continuous wave sources are expected to be hosted by $M_\star>10^{11}M_\odot$ ellipticals with low specific star formation rates (sSFRs) \citep{2026A&A...706A.115T}. Morphology-based studies also indicate that high metallicity, low sSFR, and prominent classical bulges in elliptical galaxies act as signposts for high mass and mass ratio binaries, aiding targeted searches \citep{2024ApJ...961...34B, 2026ApJ..1003..229H}.

In this paper we ask: given the inferred hardening rates in core galaxies, what are the odds that a local core galaxy hosts an SMBHB? Further, how likely is it that a local SMBHB is emitting in the pulsar timing array frequency band? To answer these questions we use the core-formation model of \cite{2026arXiv260107762H} (HGB26) as a framework for estimating the probability that a core galaxy hosts a binary as a function of observable galaxy properties.

The paper is organized as follows: In section \ref{sec:data} we describe the observed catalog of core galaxies and the simulated massive-galaxy sample used as input to our model. In section \ref{sec:methods} we outline the core-scouring model, describe our implementation of a galaxy stellar mass function, and introduce the logistic regression method used to estimate binary probabilities. In section \ref{sec:results} we present our results, including  inferred binary demographics and model performance. In section \ref{sec:highlight} we apply the model to observed core galaxies and discuss the main caveats of the analysis. We summarize our conclusions in section \ref{sec:summary}.

Throughout we adopt $\Omega_M=0.3$, $\Omega_\Lambda=0.7$, $H_0=70 \ \mathrm{km\ s^{-1}\ Mpc^{-1}}$ and ensure all values/observations conform to this cosmology. Additionally, we adopt a \cite{2003PASP..115..763C} initial mass function (IMF) and all magnitudes used are in the Vega system.

\section{Data}
\label{sec:data}
The datasets we use in our analysis include a catalog of core galaxies with derived surface brightness parameters and a selection of simulated galaxies from the output of the IllustrisTNG suite of cosmological simulations. We describe these data and their origins below.

\subsection{core galaxy Sample}

We include 116 ETGs with derived Nuker law parameters and stellar mass-to-light ratios \citep{2007ApJ...664..226L, 2009ApJS..181..486H}. The functional form of the Nuker law is given by \citep{1995AJ....110.2622L}
\begin{equation}
    I(R)=2^{\frac{(\beta-\gamma)}{\alpha}}I_b\left(\frac{R}{R_b}\right)^{-\gamma}\left[1+\left(\frac{R}{R_b}\right)^{\alpha}\right]^{\frac{(\gamma-\beta)}{\alpha}},
    \label{eq: Nuker}
\end{equation}
where $R$ is the projected radial coordinate and $R_b$ is the core or `break' radius, indicating the transition between the inner and outer profile. The parameter $\gamma$ sets the logarithmic slope of the profile as $R\rightarrow0$, while $\beta$ sets the slope as $R\rightarrow\infty$, with $\alpha$ controlling the sharpness of the transition between regimes. Lastly, $I_b=I(R_b)$ is the intensity at the break radius.

Photometric data for the galaxies in the sample were obtained from various sources. WFPC1 and WFPC2 $V$-band observations were obtained by \cite{1995AJ....110.2622L,1997AJ....114.1771F,2005AJ....129.2138L}. $R$-band WFPC2 photometry comes from \cite{2001AJ....121.2431R}, while NICMOS $H$-band photometry was obtained by \cite{2001AJ....122..653R} and \cite{2000ApJS..128...85Q}. Finally, \cite{2003AJ....125..478L} obtained WFPC2 $I$-band photometry for the brightest cluster galaxies (BCGs) in the sample. \citet{2007ApJ...664..226L} have homogenized all parameters to $V$--band values.

For each galaxy we use a set of quantities derived from observations to compare with the core-formation model. These include the V-band luminosity $L_V$, stellar mass-to-light ratio $\Upsilon_\star$, effective (half-light) radius $R_{\rm eff}$, central velocity dispersion $\sigma_\star$, and stellar mass deficit $M_{\rm def}$. We obtain the luminosities, mass-to-light ratios, and effective radii from \cite{2009ApJS..181..486H} and we take central velocity dispersion values from the HyperLeda\footnote{http://leda.univ-lyon1.fr/} database \citep{2014A&A...570A..13M}. We use the stellar mass deficits derived in HGB26 for each core galaxy in our catalog. Taken together these galaxy properties allow us to construct the observable features used in the logistic regression analysis described below.

\subsection{Simulated Cores from IllustrisTNG}

We use the $z=0$ subhalos (galaxies) and merger trees from the IllustrisTNG \citep{2018MNRAS.475..676S, 2018MNRAS.473.4077P,2019ComAC...6....2N} simulations as the input to the core-scouring model. The IllustrisTNG project is a cosmological magnetohydrodynamical simulation suite for galaxy formation and evolution, spanning a range of volume and resolution. The main components of the simulation suite we use here are the \texttt{sublink} \citep{2015MNRAS.449...49R} merger trees for tracing the merger histories of our simulated catalog. Our initial galaxy sample consists of all subhalos with total stellar masses above $10^{10}M_\odot$ in both the TNG300-1 simulation \citep{2019ComAC...6....2N} and the TNG-Cluster zoom-in simulation \citep{2024A&A...686A.157N}. We use the 300 ${\rm cMpc}^{3}$ volume with supplementary TNG-Cluster subhalos to provide us with the largest sample of massive galaxies and to span the range of galaxy masses in the observed population of cores. This is at the expense of mass and spatial resolution, however in general the IllustrisTNG suite lacks the resolution scale required to resolve core structure, which is $\lesssim0.5$ kpc for all except the largest cores, and does not explicitly evolve SMBHBs through the loss-cone scattering and gravitational-wave regimes. We therefore use the simulated galaxy properties and merger histories as inputs to a semi-analytic hardening model. This allows us to connect cosmological assembly histories of massive galaxies to predictions for core formation and the SMBHB population.

Not every subhalo above our mass threshold meets the criteria to be a core galaxy. A subhalo at the $z=0$ snapshot is selected for our simulated catalog if it is quiescent, has experienced at least one major ($q_{\rm gal}>1/3$), gas-poor merger since $z=1$ \citep{2006ApJ...652..270B, 2009ApJ...697.1369B,2012A&A...548A...7L}, and has had at least one major, gas-rich merger between $z=1$--$2$ \citep{2009ApJS..181..135H,2009ApJS..181..486H}. We consider a subhalo to be quiescent if its specific star-formation rate (${\rm sSFR}$) is less than $1/3t_H$ where $t_H$ is the Hubble time. Similarly, a merger is considered gas-poor if ${\rm sSFR}<1/3t_H(z')$ where $z'$ is the redshift of the merger and the remnant subhalo remains quiescent post-merger (i.e., there are no merger-induced starbursts). These cuts are intended to isolate massive, quiescent galaxies whose merger histories are consistent with the formation of depleted stellar cores. The selected 1151 systems form the parent population for the core-scouring calculations described in \ref{ssec:core formation}.

\begin{figure}
    \centering
    \includegraphics[width=1.0\linewidth]{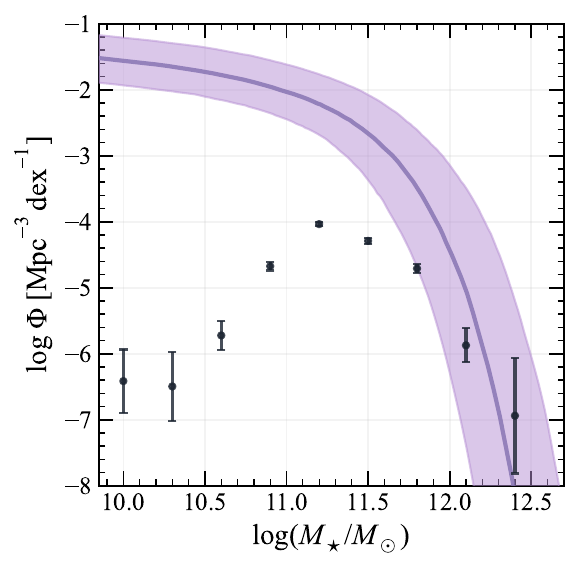}
    \caption{The purple solid curve shows the local galaxy stellar mass function used to down-sample the catalog of IllustrisTNG subhalos used in our core-formation model. It is described by a single Schechter function whose normalization, amplitude, and index controlling the sharpness of the transition from low- to high-mass are selected to be consistent with the nanohertz gravitational-wave background. The pink shaded region depicts the 16\% and 84\% confidence regions. The black data points with error bars show the mean core galaxy stellar mass function (above $10^{10}M_\odot$) produced by our model across 1000 iterations.}
    \label{fig:GSMF}
\end{figure}

The combined TNG300-1 and TNG-Cluster catalog is not volume-limited and requires down-sampling to a galaxy stellar mass function (GSMF) before any statistical analysis can be conducted. In our analysis we describe the local GSMF in terms of a single Schechter function \citep{1976ApJ...203..297S}
\begin{equation}
    \Phi(M_\star)=\ln(10)\phi_0\left(\frac{M_\star}{M_\phi}\right)^{1+\alpha_\phi}\exp\left(-\frac{M_\star}{M_\phi}\right)
\end{equation}
where $\phi_0$ is the GSMF normalization, $M_\phi$ is the characteristic mass determining the location of the 'knee', and $\alpha_\phi$ is an index controlling the sharpness of the transition from low to high mass galaxies. We fix these parameters to the maximum likelihood values determined by analysis of the North American Nanohertz Observatory for Gravitational waves (NANOGrav) 15 year datset \citep{2023ApJ...952L..37A} ($\phi_0=-2.13{\rm \ cMpc^{-3} \ dex^{-1}}$, $M_\phi=10^{11.3}M_\odot$, $\alpha_\phi=-1.21$). Figure \ref{fig:GSMF} shows the GSMF with its 16 and 84 percentile confidence intervals.

We select this GSMF for two main reasons. The first is that it is consistent with the inferred GWB by construction and the constraints from the GWB are compelling \citep{2024ApJ...971L..29L,2024PhRvD.110f3020S,2025PhRvD.111f3067L,2026ApJ...997..188M}. The second is that the alternative, electromagnetically informed, GSMF under consideration \citep{2024ApJ...971L..29L} is very similar to the NANOGrav posterior GSMF at the high mass end and therefore makes little difference in the construction of our input datasets.

\section{Methods} \label{sec:methods}
Here we briefly outline the core-scouring model used to generate simulated catalogs of core galaxies. We then introduce the logistic regression method used to estimate binary probabilities.

\subsection{Core-Scouring Model}
\label{ssec:core formation}

We use the same core-formation model described in HGB26, which we outline here. We model the stellar density profile of each merger remnant using a Dehnen profile \citep{1993MNRAS.265..250D}
\begin{equation}
    \rho_\star=\frac{(3-\lambda)M_{\rm bulge}}{4\pi}\frac{r_c}{r^\lambda(r+r_c)^{4-\lambda}},
    \label{eq: dehnen}
\end{equation}
with $M_{\rm bulge}=0.615 M_\star$ \citep{2014ApJ...788...11L, 2014MNRAS.441..599B} and $r_c=(2^{1/(3-\lambda)}-1) \ r_{1/2}$. The half-mass radius is then given by the scaling-relation \citep{2008MNRAS.386..864D}
\begin{equation}
    r_{1/2}=239\left(\frac{M_{\rm bulge}}{10^9 \ M_\odot}\right)^{0.569} \ \mathrm{[pc]}\ \ \pm \ 0.2 \mathrm{\ dex}.
\end{equation}
The logarithmic inner power-law index $\lambda$ is assigned based on the gas content of the previous merger. If the last merger was gas-rich, the remnant is initialized with a starburst-like cusp. We take the fraction of stars formed as a result of a merger-induced starburst to be
\begin{equation}
    f_\mathrm{inj}=\left[1+\left(\frac{M_\mathrm{bulge}}{M_0}\right)^{\nu}\right]^{-1}.
\end{equation}
\cite{2009ApJS..181..486H} found the best-fit values for cusp ellipticals are $(M_0, \nu)=(10^{9.2}, 0.43)$, where $M_0$ is in solar masses. We adopt these along with a constant factor of 2 intrinsic scatter at each mass. The mass in stars injected into the nucleus is then $M_{\rm inj}=f_{\rm inj} M_{\rm bulge}$. Under the assumption the nuclear starburst is contained within $r_c$ the inner logarithmic slope is
\begin{equation}
    \lambda_{\rm cusp}=\frac{\log\{8[f_{\rm inj} +2^{\lambda_0-3}(1-f_{\rm inj})]\}}{\log2}
\end{equation}
using the fact $M_\mathrm{inj}=4\pi \int_0^{r_c}[ \rho_\star(\lambda_{\rm cusp})-\rho_\star(\lambda_0)]r^2dr$. We take $\lambda_0=1$ for the pre-starburst inner power-law index.

If instead the last merger was gas-poor the slope parameter $\lambda$ is determined by the amount of mass ejected from the nucleus during core-scouring
\begin{equation}
    \lambda_{\rm core}=\frac{\log(2^{\lambda'}-8f_{\rm ej})-\log(1-f_{\rm ej})}{\log 2}
\end{equation}
where $f_{\rm ej}=M_{\rm ej}/M_{\rm bulge}$ is the fraction of mass ejected and $\lambda'$ is the inner power-law index of the primary progenitor, since light profile shapes are largely conserved in gas-poor mergers \citep{2005MNRAS.362..184B}.

We model the dark matter component with an NFW profile
\begin{equation}
    \rho_\mathrm{dm}(r)=\frac{\rho_s}{(r/r_s)(1+r/r_s)^2},
\end{equation}
where $r_s$ is the scale radius defined such that the scale density obeys $\rho_s=4 \ \rho_s(r_s)$. In general, the scale density is \begin{equation}
    \rho_s=\frac{200}{3}\rho_\mathrm{cr}(z)\frac{{C}^3}{\mathrm{ln}(1+{C})-{C}/(1+{C})},
\end{equation} 
where $\rho_\mathrm{cr}(z)=3H_0(z)^2/(8\pi G)$ is the critical density of the Universe at redshift $z$ and ${C}$ is the concentration parameter (here we adopt the prescriptions of \citealt{2016MNRAS.457.4340K}).

The remnant galaxy described by the stellar and dark components is populated with an SMBH pair of total mass $M_{\rm \bullet}=M_{\rm bh, 1}+M_{\rm bh, 2}$ and mass ratio $q_{\rm \bullet}=M_{\rm bh, 2}/M_{\rm bh, 1}\le1$ at an initial separation of $a_{\rm init} = 2r_{1/2}$ with initial eccentricity $e_0=0$. The black hole masses are assigned using the \cite{2013ARA&A..51..511K} $M_{\rm bulge}$--$M_{\rm bh}$
relation with intrinsic scatter
\begin{equation}
    M_\mathrm{BH}=0.49\times10^9\left(\frac{M_{\rm bulge}}{10^{11}M_\odot}\right)^{1.17}M_\odot \pm0.28 \ \mathrm{dex}.
\end{equation}
We apply the relation to both progenitor galaxies so that $q_{\rm \bullet}$ is determined self-consistently. For multiple mergers we conserve total black hole mass by setting the initial mass of the primary black hole accordingly.

The orbital evolution of each black hole pair is followed through dynamical friction, stellar scattering, and gravitational wave emission. We model dynamical friction using the \cite{1943ApJ....97..255C} prescription
\begin{equation}
\dot a_{\rm df}=-\tau_{\rm dyn}\frac{4\pi G^2 M_{\rm bh}\rho_{\rm gal}(r)}{ v_c^2}\ln\Lambda
\end{equation}
where $\tau_\mathrm{dyn}=a/v_c$ is the dynamical time, $v_c$ is the circular orbital speed. We apply a constant Coulomb Logarithm $\ln\Lambda=15$ consistent with the results of \cite{2017MNRAS.464.3131K} and \cite{2017MNRAS.471.4508K}. Dynamical friction becomes inefficient as the black hole pair approaches the hard binary separation
\begin{equation}
    a_h=\frac{G\mu_\bullet}{4\sigma_\star^2},
\end{equation}
such that $\mu_\bullet=M_{\rm bh,1}M_{\rm bh,2}/M_{\rm \bullet}$ and $M_\star(a_h)=2M_{\rm \bullet}$. Therefore we apply an exponential turn-off $\exp(-a_h/a)$.

Once the pair reaches the hard-binary regime its subsequent evolution is dominated by stellar scattering. The hardening rate, assuming a full loss-cone, becomes \citep{1996NewA....1...35Q, 2006ApJ...651..392S,2015MNRAS.454L..66S}
\begin{equation}
    \dot{a}_\mathrm{lc}=-\hat H H(q, a)\frac{G\rho_\star}{\sigma_\star}a^2,
    \label{eq: loss cone hardening}
\end{equation}
where $\rho_\star$ is the stellar density fixed at the black hole pair influence radius $r_{\rm inf}=r_c/[(2M_\bullet/M_{\rm bulge})^{1/(\lambda-3)}-1]$ and $H$ is the dimensionless hardening rate. We use the fits to the N-body results provided by \cite{2006ApJ...651..392S} for $H$ as a function of $q_\bullet$ and $a$. The parameter $\hat H$ is an efficiency parameter introduced by HGB26 to ensure simulated mass deficits are consistent with those in observed galaxy cores. Values of $\hat H$ below unity indicate problems with the assumption of a full loss-cone while values above unity imply additional physics not implemented in the model, most notably eccentric orbits and SMBH recoil. We discuss potential effects of these components on our analysis in section \ref{sec:highlight}. We set $\hat H=1.64$ in accordance with the results of HGB26.

The mass ejection rate is 
\begin{equation}
    \frac{dM_{\rm ej}}{da}=-\frac{J(q, a)}{\hat H}\frac{M_{\rm bin}}{a}\frac{|\dot a_{\rm lc}|}{|\dot a_{\rm df} + \dot a_{\rm lc} + \dot a_{\rm gw}|}.
\end{equation}
The factor involving the hardening rates $\dot a_i$ is an attenuation introduced to ensure that mass is ejected only as a result of hardening caused by loss-cone scattering. The dimensionless mass ejection rate is a function of the dimensionless hardening rate such that 

\begin{equation}
    J=\frac{1}{H}\int^\infty_0f(v,\sigma_\star)\frac{\sigma_\star}{v}4\pi v^2\ dv \int^\infty_0 F_{\rm ej}(x)\ dx,
\end{equation}
where $v$ is the speed of an interacting star, $x$ is the ratio of maximum to minimum impact parameters, $f(v,\sigma_\star)$ is the Maxwellian velocity distribution, and $F_{\rm ej}$ is the fraction of ejected stars. Thus $\hat H$ above unity decreases the fraction of ejected stars, leading to smaller mass deficits, all else being equal.

The total mass ejected by an SMBH pair in a single merger is
\begin{equation}
    M_{\rm ej}=\int^{a_{\rm init}}_{a_0}\frac{dM_{\rm ej}}{da}da,
\end{equation}
where $a_0$ is the final pair separation. If the pair coalesces before $z=0$ then the final separation is the innermost stable circular orbit of the primary black hole. If the pair does not coalesce then the final separation is the separation at $z=0$. The total mass deficit measured for each core galaxy is
\begin{equation}
    M_{\rm def}=\sum^{N_{\rm dry}}_{i=1}M_{\rm ej,i}
\end{equation}
where $N_{\rm dry}$ is the number of most recent, consecutive, major gas-poor mergers in the subhalo's merger history.

At subparsec separations the binary detaches from its environment and gravitational-wave emission becomes the dominate hardening mechanism. The semi-major axis of a circular orbit evolves under \citep{1964PhRv..136.1224P}
\begin{equation}
    \dot a_\mathrm{gw}=-\frac{64}{5}\frac{G^3}{c^5}\frac{M_\mathrm{\bullet}^3}{q_\bullet(1+q_\bullet)^2a^3}.
\end{equation}
and coalescence occurs rapidly.

If a black hole pair fails to coalesce before $z=0$ it is either a gravitationally unbound \emph{dual} SMBH or a bound \emph{binary} SMBH. For our purposes we define a dual (binary) SMBH at $z=0$ to have final separation greater (less) than the hard binary separation. An alternative definition using a constant value of $100$ pc as the threshold yields nearly identical results.

In the special case that a black hole pair fails to coalesce before its host galaxy undergoes a subsequent major merger we label the system as an SMBH \emph{triplet} and do not include it in the analysis. Instances of this occurring in the model are largely limited to central galaxies early in their evolution when major mergers are frequent.

% \subsection{core galaxy Stellar Mass Function}

% The sample of observed core galaxies used in HGB26 is heterogeneous and therefore does not represent a statistical sample. Similarly, the sample of simulated cores derived from the IllustrisTNG simulations is not volume-limited because it contains subhalos from TNG300-1 and TNG-Cluster. To mitigate this and allow for a meaningful statistical analysis of model outputs we downsample the model inputs using GSMFs.

% We create a total GSMF for the combined TNG300-1 and TNG-Cluster sample by combining the individual GSMFs where they smoothly intersect. We also conduct our analysis using the observationally motivated GSMFs published in Liepold and Ma (2024), Agazie et al. (2023) and Leja (2020) (Fig. 1).

\subsection{Logistic Regression}

Our goal is to determine the probability of a galaxy with given properties to host an SMBHB in the local Universe. In our model a galaxy hosts a single SMBH at $z=0$, a binary SMBH, or a dual SMBH. Observationally, a galaxy hosting a dual SMBH will either be a cusp or a core galaxy. If the dual SMBH is the result of the first major, gas-poor merger the galaxy has had in its recent merger history then the mass deficit will be so small as to be unobservable, as the pair by definition has not entered the loss-cone scattering phase of its evolution. Conversely, if an observable core does exist within the host galaxy it will have been scoured out by a previous binary. In the latter case a core galaxy hosting a dual would be observationally identical to one which has only a single SMBH at its center. As a result we can say that the state single/binary is a label (or target) for the simulated core. The state single or binary is ultimately a function of galaxy properties (or features).

Logistic regression operates in odds space, rather than probability space, so we first convert via the ${\rm logit}$ function such that
\begin{equation}
    \ln{\cal O}={\rm logit}({\cal P})\equiv\ln \left( \frac{{\cal P}}{1-{\cal P}}\right)
\end{equation}
where ${\cal O}$ is the odds and ${\cal P}$ is the probability. In general, logistic regression produces a linear function
\begin{equation}
    \ln{\cal O}=\beta\vec X
\end{equation}
where $\vec X$ are the independent variables (plus intercept) and $\beta$ is a diagonal matrix containing the coefficients.

For our analysis we use three physically motivated features, each with low variance-inflation factors (${\rm VIF}<2$ always). The first feature we expect to be an important predictor of binary presence is the ratio of the stellar mass deficit to the total stellar mass $\log(M_{\rm def}/M_\star)$. This follows because a binary at parsec scale separations will not have ejected as much mass at fixed total stellar mass as a coalesced binary.

The second feature we use is the quantity $\log(\rho_\star/\sigma_\star)$ which is a scaled version of the traditional hardening parameter $s=HG\rho_\star/\sigma_\star=\dot a_{\rm lc}/a^2$.

The third feature we include is the compactness of the bulge $\log (R_{\rm eff}/M_{\rm \star})$. Larger effective radii for the same stellar mass will result in black hole pairs that have a larger range of separations to cover with fewer stars to interact with. For the simulated core galaxies we take $R_{\rm eff}=0.75r_{1/2}$ \citep{1993MNRAS.265..250D}.

\begin{figure*}
    \centering
    \includegraphics[width=0.90\linewidth]{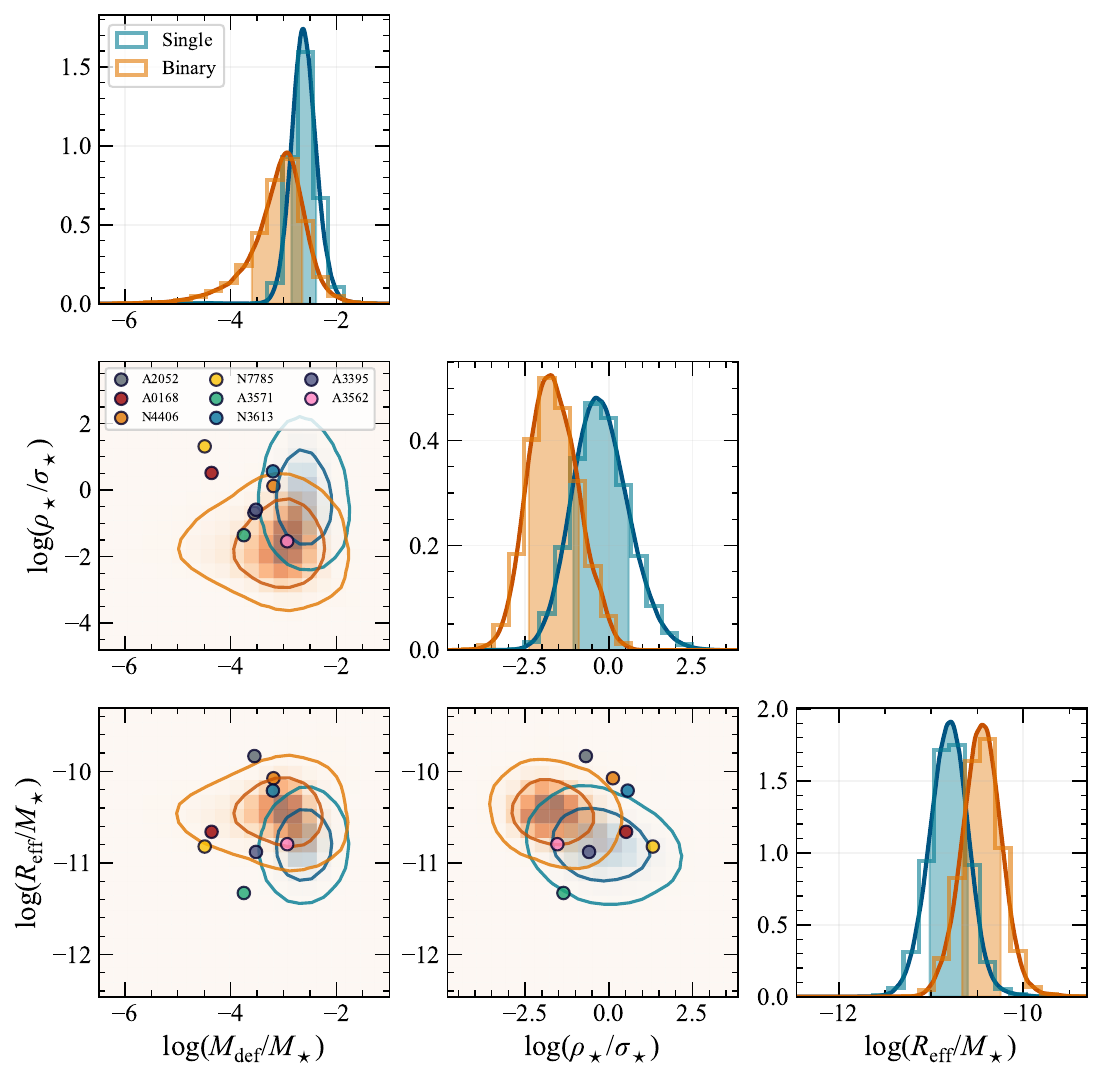}
    \caption{We show the feature space distributions according to single (blue) or binary (orange) occupation. The histograms show the binned posteriors, while the smooth curves are kernel density estimations. The shaded region shows the 16\% and 84\% confidence region. Colored points represent the eight model-selected candidates described in section \ref{application}. The feature distributions are well separated, indicating that they are informative for the logistic classifier. The lack of significant covariance between features reflects low variance-inflation factors and  allows the derived regression coefficients to be interpreted physically.}
    \label{fig:feature corner}
\end{figure*}

With three features the log-odds becomes
\begin{equation}
    \ln{\cal O}={\cal A}\log\frac{M_{\rm def}}{M_\star} + {\cal B}\log\frac{\rho_\star}{\sigma_\star}+{\cal C}\log\frac{R_{\rm eff}}{M_\star}+{\cal D}
\label{eq:log-odds}
\end{equation}
where ${\cal A}, {\cal B},$ and ${\cal C}$ are coefficients with magnitudes corresponding to the predictive strengths of the features, as long as the features are properly scaled. The coefficient ${\cal D}$ is a normalization, corresponding to the case when each of the features is identically zero. We perform standard scaling for each feature, subtracting the mean and dividing by the standard deviation. For this analysis we use units of $M_\odot$ for the stellar mass and mass deficit, ${\rm km/s}$ for the velocity dispersion, $M_\odot/{\rm pc}^{3}$ for the central density, and ${\rm kpc}$ for the effective radius. We show the feature distributions and credible intervals for both binary and single SMBHs in Figure \ref{fig:feature corner}.

\section{Results}
\label{sec:results}

\subsection{Binary Host Odds Estimation}
\begin{figure*}
    \centering
    \includegraphics[width=0.90\linewidth]{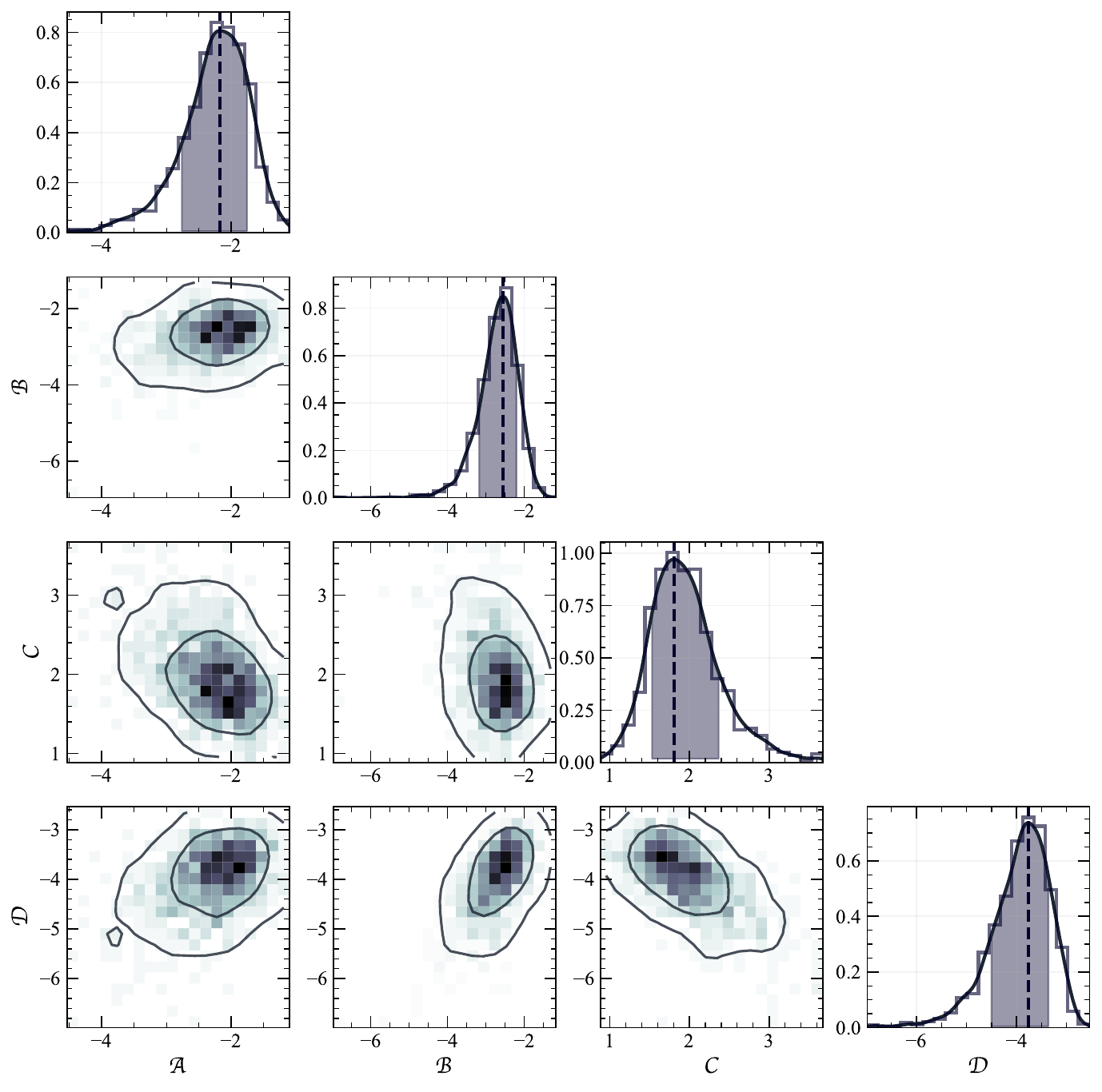}
    \caption{We show the logistic regression coefficient posteriors for 1000 iterations of our core-scouring model. The distributions are well constrained, capturing robust information about host properties that result in surviving binaries. The histograms show the binned posteriors, while the smooth curves are kernel density estimations. The shaded region shows the 16\% and 84\% confidence region, while the dashed lines show the maximum-likelihood estimate for each coefficient. The 2D histograms show the covariances for each coefficient, which appear as expected for low variance-inflation factors among the features. We find the maximum likelihood values of the coefficients to be ${\cal A}=-2.17, \ {\cal B}=-2.54, \ {\cal C}=1.80,$ and ${\cal D}=-3.77$.}
    \label{fig:coefficients}
\end{figure*}

Figure \ref{fig:coefficients} shows the posterior distribution of the logistic regression coefficients for 1000 iterations of the core-scouring model. The coefficient distributions are well constrained and broadly consistent across realizations, indicating the features capture robust information about binary occupation in the model population. We find the maximum likelihood values of the coefficients to be ${\cal A}=-2.17, \ {\cal B}=-2.54, \ {\cal C}=1.80, {\cal D}=-3.77$ (Fig. \ref{fig:coefficients}) and the models achieve a pseudo-$R^2$ of $0.51\pm0.03$. The descaled values to be used in Equation \ref{eq:log-odds} are ${\cal A}=-5.67, \ {\cal B}=-2.63, \ {\cal C}=7.20, {\cal D}=57.4$.

Using a probability threshold of 50\% to classify galaxies as hosting a binary or single SMBH, the models obtain the confusion matrix
\begin{equation}
    \begin{bmatrix}
        {\rm TN} & {\rm FP}\\
        {\rm FN} & {\rm TP}
    \end{bmatrix} = \begin{bmatrix}
    299\pm13 & 5\pm2 \\
    12\pm3 & 59\pm 7
    \end{bmatrix}  , 
\end{equation}
where the true negatives, false positives, false negatives, and true positives quoted are the median values along with their standard deviations. The resulting median classification metrics and their standard deviations are $0.95\pm0.01$ for accuracy, $0.92 \pm0.03$ for precision, and $0.82\pm0.04$ for recall. The values indicate that the model performs well at separating galaxies hosting binaries from those hosting single black holes. The high precision suggests that galaxies assigned high binary probabilities are usually true binary hosts in the model, while the slightly lower recall indicates that some binary hosts occupy regions of feature space that overlap with single SMBH systems.

The logistic regression models identify clear trends between galaxy structure and binary occupation. Galaxies with low values of $\log (M_{\rm def}/{M_\star})$ and $\log(\rho_\star/\sigma_\star)$ and high values of $\log (R_{\rm eff}/M_{\rm \star})$ are more likely to host surviving binaries. These trends are consistent with expectations from the core-scouring model. A large stellar mass deficit indicates that a binary has already ejected substantial mass and may therefore have had sufficient time to harden and coalesce. Conversely, a low mass deficit at fixed stellar mass can indicate a system in which the binary has not yet fully processed the nuclear stellar population. Similarly, low central densities imply less efficient stellar hardening, increasing the probability that a binary remains. Finally, a larger effective radius at fixed stellar mass corresponds to a more extended and diffuse galaxy in which binary evolution proceeds more slowly.

\subsection{core galaxy and Black Hole Pair Demographics}

Our resampled simulated population allows us to estimate the demographics of core galaxies and their central black hole pairs. Using the adopted GSMF we find the fraction of galaxies hosting depleted cores with stellar masses above $10^{10}M_\odot$ to be $f_{\rm core}\approx0.002$ (see Fig. \ref{fig:GSMF}). Among core galaxies the binary fraction is $f_{\rm bin}=0.18\pm0.02$. Thus while core galaxies are intrinsically rare among all galaxies a non-negligible fraction of them are expected to host surviving SMBHBs.
\begin{figure}
    \centering
    \includegraphics[width=1.0\linewidth]{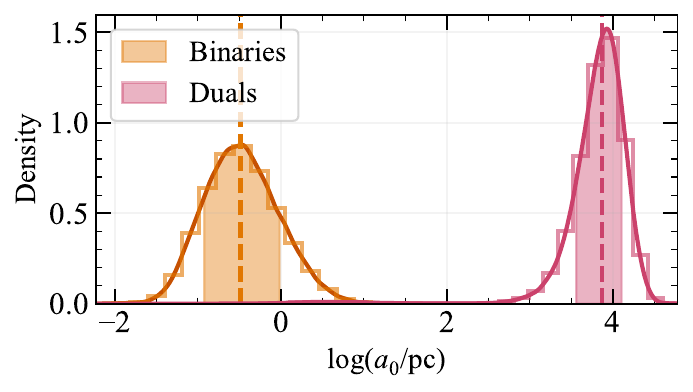}
    \caption{The predicted separation distribution for dual (pink) and binary (orange) SMBHs in local core galaxies. The histograms show the binned $z=0$ separations of each black hole pair across 1000 iterations of our core-scouring model. The smooth curves are kernel density estimations. The shaded regions show the 16\% and 84\% intervals and the dashed lines show the maximum-likelihood estimates. Duals in the model tend to cluster around the initial separation, pointing to low stellar densities and inefficient dynamical friction in the host galaxies. Binary separations peak at subparsec distances. At larger separations ($\sim10^1$--$10^3$ pc) hardening is efficient and pairs spend little time in this regime.}
    \label{fig:pair separations}
\end{figure}
The predicted separation distributions for duals and binaries are shown in Figure \ref{fig:pair separations}. Dual black holes tend to cluster around the initial separation of the SMBH pairs, indicating they are the result of inefficient dynamical friction acting at the ${\rm kpc}$ scale. Binary separations peak below 1 $ {\rm pc}$. This separation scale is particularly important because binaries at subparsec separations are close enough for gravitational wave emission to become relevant, but may still be difficult to resolve electromagnetically.
\begin{figure*}
    \centering
    \includegraphics[width=0.9\linewidth]{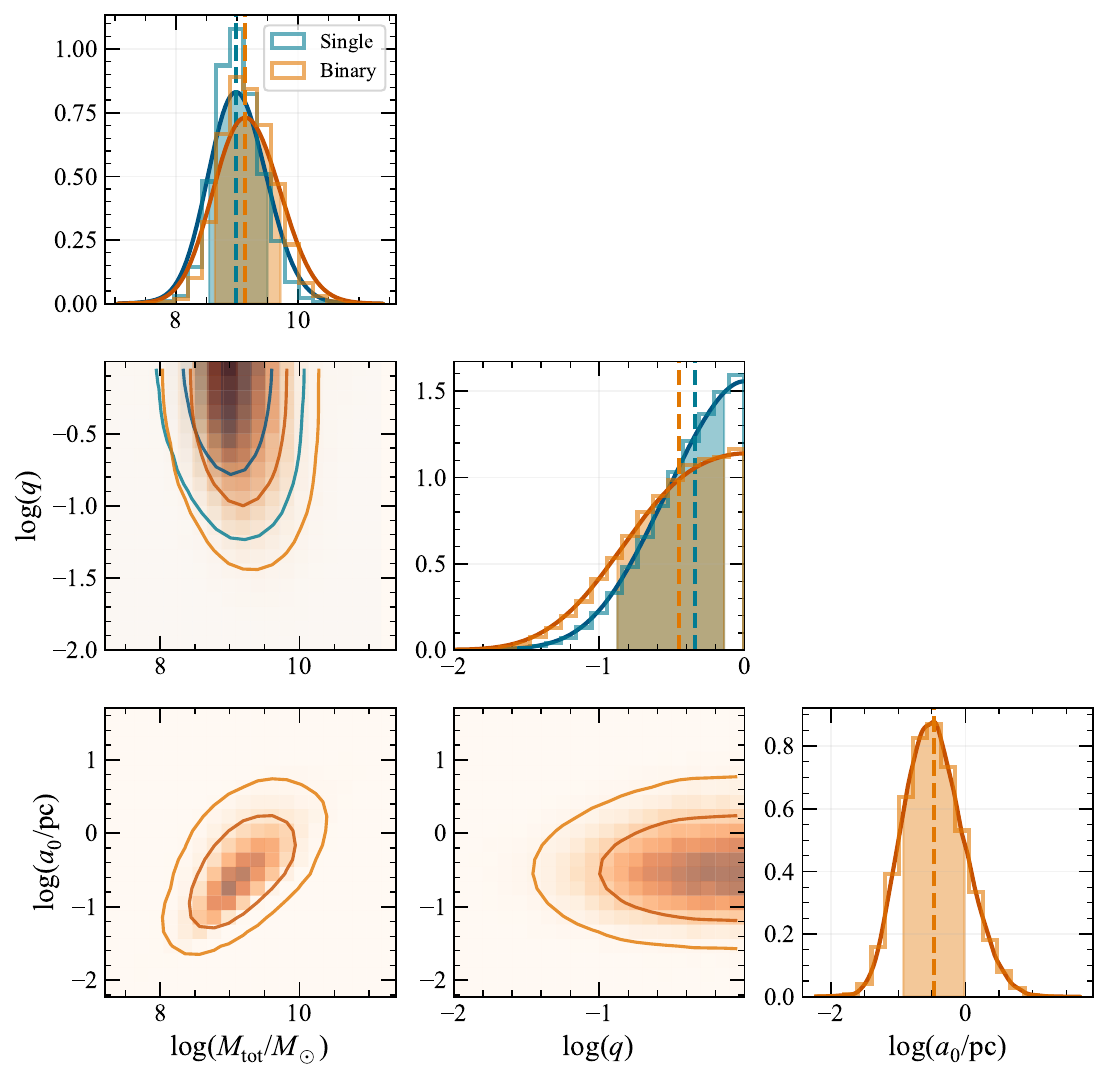}
    \caption{The distributions of total black hole mass, mass ratio, and present-day separation for binaries across all model realizations (orange) alongside distributions of black hole mass and mass ratio for single SMBHs at $z=0$ (blue). Histograms show the binned data while smooth curves are kernel density estimations. The dashed lines show the maximum-likelihood estimates while shaded regions show the 16 and 84 percentiles. Surviving binaries tend to have higher total mass and lower mass ratio than those which have coalesced, reflecting assembly via hierarchical merging. Present-day binaries with higher total mass tend to exist at larger separations.}
    \label{fig:pair distributions}
\end{figure*}
The surviving binaries are also skewed toward higher total masses and lower mass ratios relative to the binaries that have coalesced by $z=0$, as shown in Figure \ref{fig:pair distributions}. This trend is expected because more massive galaxies undergo more mergers and host more massive black holes, increasing the likelihood a binary will survive until $z=0$. The skew toward lower mass ratios reflects the hierarchical assembly of massive galaxies, in which earlier mergers are more likely to have $q_{\rm gal}$ closer to unity than later mergers.
\begin{figure}
    \centering
    \includegraphics[width=1.0\linewidth]{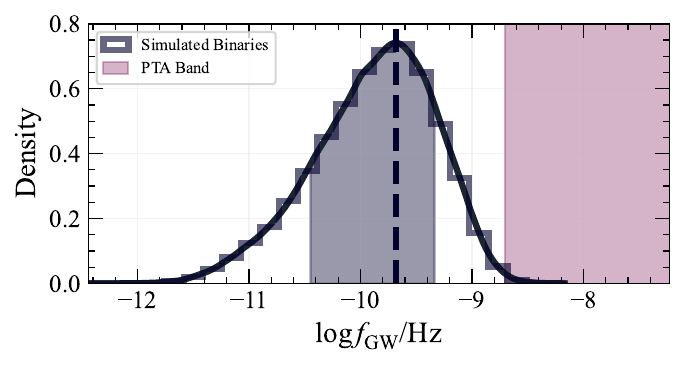}
    \caption{The histogram shows the posterior distribution of circular gravitational wave frequencies of surviving binaries across all model outputs. The smooth curve is a kernel density estimate with the purple shaded region indicating the 16 to 84 percentile range. The dashed line shows the most likely frequency ($2.1\times10^{-10} \ {\rm Hz}$). The pink shaded region shows the frequency space pulsar timing arrays are sensitive to. Only a small portion of all $z=0$ binaries in the model (0.3\%) are active as pulsar timing array sources. Most binaries are either too widely separated or else have already evolved past the relevant regime.}
    \label{fig:PTA}
\end{figure}
Lastly, we examine the gravitational-wave frequencies of surviving binaries. The distribution is shown in Figure \ref{fig:PTA}. Only a small fraction of binaries emit in the pulsar timing array frequency band. Across all 1000 model realizations we find that approximately 0.3\% of binaries lie in the PTA band. This result implies that while core galaxies may be useful for identifying nearby massive binaries, only a subset of those systems are expected to be individually relevant as PTA continuous-wave sources.

\section{Discussion} \label{sec:highlight}
We can use the results of performing logistic regression on our model outputs to make comments on the probability of a local core galaxy hosting a binary. We additionally discuss how model results impact the search for continuous-wave sources in PTA data. Lastly we comment on the impact of model assumptions on the validity of our results.

\subsection{Application to Observed core galaxies}
\label{application}
\begin{figure*}
    \centering
    \includegraphics[width=1.0\linewidth]{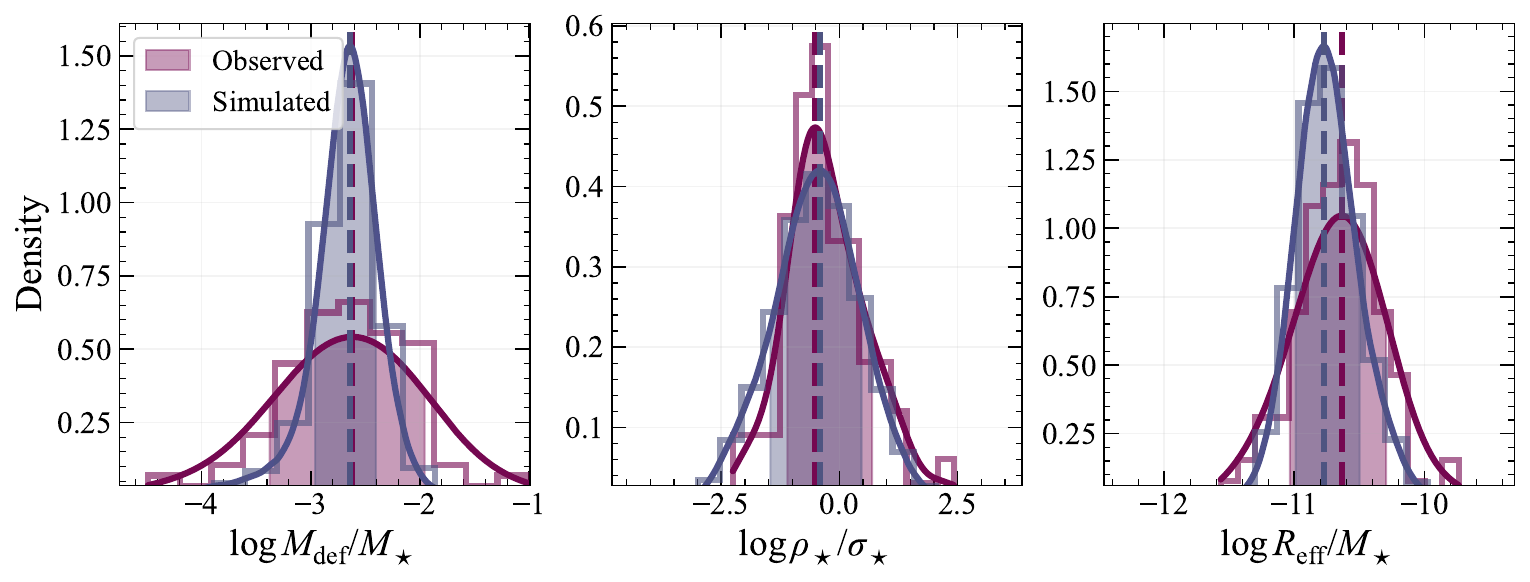}
    \caption{Comparison of the feature distributions for both the observed sample of core galaxies (pink) and all cores formed in our models (purple). Histograms show the binned parameters while smooth curves are kernel density estimations. Shaded regions are the 16 and 84 percentile ranges and dashed lines show the maximum likelihood feature values. The observed and simulated distributions are broadly consistent--a manifestation of the mass fundamental plane and the fact the core model used in the analysis was calibrated to the mass deficits of the observed sample. Of note, however, is the tendency of the observed sample toward more diffuse systems. This is caused by the over-representation of BCGs in the observed catalog of cores. We also point out that the variance in the mass deficit at each stellar mass is much higher in the observed sample than in the model population. We discuss potential reasons for this in section \ref{ssec: impact}.}
    \label{fig:feature distributions}
\end{figure*}
We apply the logistic regression models to the catalog of observed core galaxies. For each observed galaxy we compute the three features of interest. The stellar masses, effective radii, central velocity dispersions, and stellar mass deficits are available for 99 galaxies in the observed sample. The central density, however, is not directly tabulated for every object so we estimate it from the Nuker law parameters.

To estimate the central density we use the break radius $r_b$ as a proxy for the black hole influence radius $r_{\rm inf}$. This is motivated by the tight, nearly one-to-one scaling relation between the break radius and the black hole influence radius \citep{2016Natur.532..340T}
\begin{equation}
    \log\frac{r_{\rm inf}}{{\rm kpc}}=(-0.18\pm 0.21)+(1.00\pm0.09)\log \frac{r_b}{{\rm kpc}}
    \label{eq:rb rinf}
\end{equation}
with an intrinsic scatter of $0.16 \ {\rm dex}$. The stellar mass interior to the break radius is
\begin{equation}
    M_\star{(<r_b)=\pi I_br_b^2\Upsilon_\star}.
    \label{eq:core mass}
\end{equation}
Combining equations \ref{eq:rb rinf} and \ref{eq:core mass} gives us an estimate for the core density:
\begin{equation}
    \rho_\star=\frac{3I_b}{4r_b}\Upsilon_\star
\end{equation}
Although approximate, this procedure enables a consistent comparison between the observed and simulated core galaxies. Figure \ref{fig:feature distributions} shows a comparison between the feature distributions for the observed and simulated core galaxies. The distributions are broadly consistent, as expected from the mass fundamental plane and the fact the core-scouring model implemented here is calibrated to the mass deficits of the observed sample. Of note, however, is the observed sample's skew toward less compact galaxies. This is caused by the fact the observed catalog contains an over-representation of BCGs. We will return to this point in section \ref{ssec: impact}.

For each core galaxy we evaluate the binary probability using all 1000 logistic regression realizations and report the median probability, together with the 16th and 84th percentiles. We find eight observed galaxies with estimated binary probabilities above 50\% (see Table \ref{tab:probs}). Each of the candidates are promising targets for follow-up as they combine observable core structure with model-inferred conditions favorable to stalled or slowly evolving binaries. In particular, high-probability systems tend to occupy regions of feature space associated with low central hardening efficiency and incomplete core scouring (see Fig. \ref{fig:feature corner}). Such galaxies may retain binaries at separations relevant for electromagnetic searches.
% Five of these galaxies are BCGs located at distances greater than 100 Mpc. M86 stands out as the most promising candidate, with a high estimated binary probability and a relatively small distance.

% VLA observations of M86 reveal it to be a faint radio source with point-like morphology, lacking radio-mode AGN feedback \citep{1985ApJ...291...32B, 2009AJ....138.1990C, 2010MNRAS.404..180D, 2022ApJS..258...30G}. It is the second brightest X-ray elliptical in Virgo, with NGC 4486 (M87) the brightest, likely due to supersonic motion through the intracluster medium \citep{2019ApJS..241...36K}. M86 also shows signs of galaxy--galaxy interaction, most notably with the neighboring spiral galaxy NGC 4438 \citep{2012ApJ...748..123S,2018MNRAS.481.4472L}.
% 
\begin{deluxetable*}{lccccccc}
% \tabletypesize{\scriptsize}
\setlength{\tabcolsep}{3pt}
\tablecaption{High-Probability Binary Host Candidates\label{tab:probs}}
\tablewidth{0pt}
\tablehead{
\colhead{Designation} &
\colhead{Distance} &
\colhead{$\log(M_\star/M_\odot)$} &
\colhead{$\sigma_\star$} &
\colhead{$R_{\rm eff}$} &
\colhead{$\log(M_{\rm def}/M_\odot)$} &
\colhead{$\log[\rho_\star/(M_\odot\,{\rm pc}^{-3})]$} &
\colhead{Estimated Probability} \\
\colhead{} &
\colhead{(Mpc)} &
\colhead{} &
\colhead{(${\rm km \ s^{-1}}$)} &
\colhead{(kpc)} &
\colhead{} &
\colhead{} &
\colhead{(\%)}
}
\startdata
A2052-BCG    & 150.9 & 12.11 & 218.5 & 189.2 & 8.56 & 1.65 & $99.98^{+0.02}_{-0.08}$ \\
A0168-BCG    & 177.9 & 12.16 & 346.4 & 31.62 & 7.80 & 3.05 & $97.85^{+1.85}_{-9.02}$ \\
NGC 4406 (M86) & 17.9 & 11.36 & 231.4 & 19.27 & 8.17 & 2.49 & $92.12^{+5.14}_{-12.13}$ \\
NGC 7785     & 49.9  & 11.48 & 252.3 & 4.56  & 6.99 & 3.72 & $76.16^{+20.47}_{-42.61}$ \\
A3571-BCG    & 151.3 & 13.40 & 317.7 & 117.9 & 9.65 & 1.15 & $58.46^{+24.84}_{-30.47}$ \\
NGC 3613     & 30.8  & 11.01 & 212.5 & 6.31  & 7.81 & 2.89 & $58.35^{+21.03}_{-25.63}$ \\
A3395-BCG    & 186.1 & 12.79 & 271.7 & 81.28 & 9.27 & 1.84 & $55.68^{+19.95}_{-19.80}$ \\
A3562-BCG    & 185.6 & 12.84 & 307.4 & 110.8 & 9.91 & 0.95 & $50.25^{+10.07}_{-10.15}$ \\
\enddata
\tablecomments{Uncertainties on the estimated probabilities represent
the reported upper and lower intervals.}
\end{deluxetable*}
\begin{deluxetable}{lccc}
\tablecaption{Predicted Binary Properties\label{tab:predicted}}
\tablewidth{0pt}
\tablehead{
\colhead{Designation} &
\colhead{$\log(M_\bullet/M_\odot)$} &
\colhead{$q$} &
\colhead{$a_0$} \\
\colhead{} &
\colhead{} &
\colhead{} &
\colhead{(pc)}
}
\startdata
A2052-BCG      & $8.25^{+0.26}_{-0.25}$ & $0.29^{+0.45}_{-0.13}$ & $0.14^{+0.15}_{-0.08}$ \\
A0168-BCG      & $8.70^{+0.22}_{-0.30}$ & $0.38^{+0.32}_{-0.23}$ & $0.55^{+0.60}_{-0.28}$ \\
NGC 4406 (M86) & $8.39^{+0.21}_{-0.24}$ & $0.44^{+0.36}_{-0.23}$ & $0.06^{+0.07}_{-0.03}$ \\
NGC 7785       & $8.58^{+0.01}_{-0.05}$ & $0.53^{+0.05}_{-0.10}$ & $0.55^{+0.31}_{-0.03}$ \\
A3571-BCG      & $9.80^{+0.10}_{-0.07}$ & $0.34^{+0.16}_{-0.20}$ & $3.39^{+0.24}_{-2.03}$ \\
NGC 3613       & $8.39^{+0.21}_{-0.20}$ & $0.43^{+0.34}_{-0.24}$ & $0.06^{+0.09}_{-0.03}$ \\
A3395-BCG      & $9.04^{+0.30}_{-0.27}$ & $0.34^{+0.40}_{-0.21}$ & $0.50^{+0.61}_{-0.29}$ \\
A3562-BCG      & $9.43^{+0.24}_{-0.28}$ & $0.38^{+0.36}_{-0.23}$ & $0.31^{+0.32}_{-0.15}$ \\
\enddata
\tablecomments{$M_\bullet$ is the predicted binary black hole mass,
$q$ is the binary mass ratio, and $a_0$ is the predicted separation.}
\end{deluxetable}

In the following subsections we discuss known observational properties of the candidate galaxies along with connections to SMBHB activity or lack thereof.
\subsubsection{Abell 2052-BCG}
\label{2052}
The eponymous central dominant galaxy of the cool-core cluster Abell 2052, and host of the powerful radio source 3C 317. Very Long Baseline Array (VLBA) imaging at 1.7, 4.9, and 8.3 GHz revealed a compact core with bipolar outflows, along with an amorphous halo with unusually steep emission \citep{1993ApJ...416...51Z}. \citet{2004A&A...422..515V} estimated a radiative age of only $\sim 170$ years for the inner $\sim 10$ parsecs; extended radio plasma emissions appear older, suggesting Abell 2052-BCG is a restarted radio galaxy. Chandra and ROSAT imaging additionally supports the idea of repeated AGN outbursts, with evidence of past jet activity \citep{2000AJ....119...21R, 2001ApJ...558L..15B, 2003ApJ...585..227B, 2010A&A...523A..81D}.

The asymmetric nature of the gas kinematics in the nucleus of Abell 2052-BCG render a secure SMBH mass measurement using the central $[{\rm N\ II}]\lambda6583$ emission line width impossible, though \citet{2009ApJ...690..537D} use HST spectroscopy to determine an upper limit of $M_\bullet \lesssim 4.6\times 10^{9} M_\odot$. This is consistent with scaling relations \citep{2009ApJ...698..198G, 2013ARA&A..51..511K, 2013ApJ...764..184M} and the total black hole mass estimated in this work, assuming binary SMBHB occupancy. Further, VLBA observations resolving the radio core down to $\sim 3 \ {\rm pc}$ do not rule out the sub-parsec separation predicted here (Table \ref{tab:predicted}).
\subsubsection{Abell 0168-BCG}
\label{0168}
Abell 0168-BCG (UGC 00797) is the central dominant galaxy of the X-ray, double-peaked cluster Abell 0168, with offsets between optical and X-ray centers for the BCG consistent with a roughly equal-mass cluster merger \citep{2004ApJ...600..141Y}.

HST imaging of Abell 0168-BCG 
reveals one of the smallest core sizes ($r_b<0^{\prime\prime} .15$ in I-band) among cored elliptical galaxies, while at the same time showing a pronounced turnover in the surface brightness profile \citep{2003AJ....125..478L}. Figure \ref{fig:feature corner} shows this is the dominant driver of the high binary occupancy probability for Abell 0168-BCG (Table \ref{tab:probs}).
\subsubsection{NGC 4406}
\label{4406}

NGC 4406 (M86) is the second brightest X-ray source in the Virgo Cluster behind NGC 4486 (M87). Very Large Array (VLA) observations reveal a faint, point-like radio emission in the center of this giant elliptical \citep{2022ApJS..258...30G}. Star formation could dominant the observed radio emission, as NGC 4406 lacks observable signatures of radio-mode AGN feedback.

In the optical and near-infrared, NGC 4406 demonstrates a nearly flat central core with a depression ($\gamma=-0.04$), flattening of the color gradient for $r<r_b=0^{\prime\prime}.92$, and its inner isophotes have undefined position angles and ellipticities \citep{2002AJ....124.1975L,2005AJ....129.2138L}. NGC 4406 exhibits a ring of reduced surface brightness interior to $r_b$ \citep{1997ApJ...481..710C}. \citet{2002AJ....124.1975L} take this depression to be representative of the underlying stellar population, rather than attributing it to dust. One interpretation of the feature is that NGC 4406 began with a normal core and that some of the stars in the region interior to the core break radius were subsequently ejected. Additional surface brightness analysis by \citet{2009ApJS..182..216K} show NGC 4406 to have overlapping isophotes with adjacent galaxies in the Virgo Cluster, rendering the outermost profile uncertain. This and/or tidal distortions from close encounters may be the cause of significant deviation from the standard Sérsic profile at large radii and the large value of the corresponding Sérsic index $n\approx10$.

NGC 4406 is a slow rotator with a large scale (kpc) kinematically decoupled core (KDC) \citep{2011MNRAS.414..888E, 2011MNRAS.414.2923K}, a feature long associated with recent mergers \citep{2015ApJ...802L...3T}. \citet{2026ApJ..1003..229H} performed a linear discriminant analysis (LDA) on massive ETGs from the MASSIVE \citep{2014ApJ...795..158M}, ATLAS$^{\rm 3D}$ \citep{2011MNRAS.413..813C} and CALIFA \citep{2012A&A...538A...8S, 2014A&A...569A...1W} surveys to determine a ranked list of galaxies potentially hosting SMBHBs emitting in the PTA band. Their LDA model was trained on output data from the \texttt{ROMULUS25} cosmological simulations \citep{2017MNRAS.470.1121T}, informed by galaxy stellar kinematic properties. Specifically, the key predictors the study used are $\lambda_{R_e}$ the spin angular momentum of the galaxy measured at $R_e$ \citep{2007MNRAS.379..401E}, and $\Delta{\rm PA}$ the difference between the photometric position angle as measured in the galaxy image and the kinematic position angle as measured in the galaxy stellar kinematic map. \cite{2026ApJ..1003..229H} rank M86 as number 10 in their list with a normalized LDA score of 0.93, though they caution a high score can also indicate a recently coalesced ($\lesssim 1 \ {\rm Gyr})$ or wide-separation system.

Unfortunately, the existing kinematic data do not resolve the region where binary-induced perturbations are expected to be strongest \citep{2010MNRAS.407.1497M, 2014ApJ...782...39T}, prohibiting a definitive test of the binary hypothesis. High-resolution integral-field spectroscopy of M86's nucleus using the James Webb Space Telescope's NIRSpec instrument or MUSE on the Very Large Telescope would provide the detailed kinematic maps and nonparametric line-of-sight velocity distributions needed to perform the phase-space experiment.
\subsubsection{NGC 7785}
\label{7785}
 In contrast to the previous galaxies, NGC 7785 is a relatively isolated elliptical \citep{2004AJ....127.1336S} hosting a weak/radiatively inefficient AGN with a hot gas halo extending roughly 32 kpc with mass $M_{\rm gas}\simeq4.6\times 10^{9}M_\odot$ and a dark matter component consistent with those found in spirals of comparable luminosity \citep{2009A&A...497..359M, 2010ApJ...715L...1M, 2011A&A...534A..50M}.

 NGC 7785 has a small core size ($r_b=0^{\prime\prime}.07$ in V-band) \citep{2007ApJ...664..226L}. As a result, NGC 7785 has the smallest estimated stellar mass deficit in our sample ($\log(M_{\rm def}/M_\odot)=6.99$). Figure \ref{fig:feature corner} shows this is the main driver behind our binary classifier suggesting NGC 7785 as a promising host candidate.
\subsubsection{Abell 3571-BCG}
\label{3571}
The central dominant galaxy at the center of Abell 3571. The cluster itself is a weak cool-core system with temperature and optical kinematics indicative of a complex merger history \citep{2005AdSpR..36..643H}.

Decomposition using Core-Sérsic profiles \citep{2003AJ....125.2951G} produces a tension in the measurement of the core radius when compared to estimates obtained using the Nuker law \citep{1995AJ....110.2622L}. \citet{2003AJ....125..478L} measure an I-band break radius of $1^{\prime\prime}.15$ while \citet{2014MNRAS.444.2700D} measure $1^{\prime\prime}.70$ for the same band, differing by more than 25\%. Using the data from \citet{2003AJ....125..478L} and \citet{2007ApJ...664..226L} we estimate a stellar mass deficit of $\log(M_{\rm def}/M_\odot)=9.65$, lower than the typical value for cores of the same stellar mass. Conversely, \citep{2019ApJ...886...80D} using methods/results described in \citet{2013AJ....146..160R,2014MNRAS.444.2700D,2016Natur.532..340T} determine a mass deficit of $\log(M_{\rm def}/M_\odot)=10.25$, corresponding to a central SMBH mass of $9.31\lesssim \log(M_\bullet/M_\odot)\lesssim10.44$. Since one of the three features our binary classifier operates on is $\log(M_{\rm def}/M_\star)$, our estimated probability for Abell 3571-BCG depends on which surface brightness model accurately captures the core size. In this work, we have adopted the Nuker law, as the fits to the inner region are less sensitive to the global light profile than are fits using the core-Sérsic formalism \citep{2007ApJ...662..808L}. If the core of Abell 3571-BCG is truly larger the binary hypothesis becomes less likely and the core would need to be explained in terms of either an ultramassive black hole (UMBH) or else a core-enhancement via SMBH recoil post-coalescence \citep{2004ApJ...613L..37B}.
\subsubsection{NGC 3613}
\label{3613}
A luminous, highly flattened elliptical at about 30 Mpc, usually classified near E6 and regarded as the dominant galaxy of a poor group. The galaxy surface-brightness profile requires multiple components and contains visible substructure, evidence for a nontrivial assembly history \citep{2020MNRAS.492.4313D}.
\subsubsection{Abell 3395-BCG}
\label{3395}
Part of the interacting Abell 3391–3395 complex, with multiple X-ray peaks and a connecting filamentary environment. X-ray studies find temperature and density structure consistent with ongoing assembly. Observations found extended radio emission associated with Abell 3395-BCG, extending on the order of $3^{\prime}$, or roughly 170 kpc.
\subsubsection{Abell 3562-BCG}
\label{3562}
Dominant galaxy of Abell 3562, part of the A3558–A3562 complex in the core of the Shapley Concentration. The cluster has strong evidence for merger-driven dynamics. Very faint/quiet at radio wavelengths.
\subsection{Implications for PTA Continuous-Wave Searches}
The connection between core galaxies and PTA individual sources is indirect, but important. The nanohertz GWB is produced by the integrated cosmic population of SMBHBs, with the signal dominated by the most massive and nearby systems \citep{2017MNRAS.464.3131K,2017MNRAS.471.4508K,2023ApJ...952L..37A}. core galaxies host the most massive black holes in the Universe and therefore represent a natural target population for indivdual continous-wave searches.

However, our results indicate that only a small fraction of surviving binaries lie in the PTA band at a given time. Although the core galaxy binary fraction is significant, most binaries are either at separations too large to emit at PTA frequencies or have already evolved beyond the relevant stage (see Figure \ref{fig:pair separations}). Thus a high probability of hosting a binary alone is not sufficient to identify a PTA source.

The galaxies identified in Table \ref{tab:probs} therefore provide a useful but not complete target list. They should be interpreted as systems in which the presence of a surviving binary is plausible given the core-scouring model, rather than as high-probability PTA-band sources.
\subsection{Impact of Model Assumptions}
\label{ssec: impact}
Several caveats affect the interpretation and validity of our results.

First, our fiducial model assumes circular binary orbits. In reality, black hole pairs almost certainly have nonzero initial eccentricities. In fact, high initial eccentricities and eccentricity evolution are common in hydrodynamical simulations of galaxy mergers and N-body experiments \citep{2023MNRAS.526.2688R, 2025A&A...703A..86F,2026MNRAS.546ag026G}. Nonzero orbital eccentricities can significantly alter the amount of stellar mass ejected during hardening as well as the gravitational-wave inspiral timescale. Indeed, this effect could be a reason why the current model does not show as much variance as the observed $\log(M_{\rm def}/M_\star)$ distribution. The adopted scattering efficiency $\hat H=1.64$ likely absorbs some of these uncertainties, but eccentricity remains an important physical component to implement in future work.

Second, we neglect gravitational-wave recoil. During the final stages of coalescence GW emission occurs anisotropically, resulting in a kick to the remnant SMBH, with the kick velocities reaching $\sim4000 \ {\rm km/s}$ in the most extreme case \citep{1973ApJ...183..657B, 2007ApJ...659L...5C}. Numerous studies have shown that recoiled SMBHs enhance core size beyond that induced by binary hardening alone by factors of 2--3 \citep{2004ApJ...613L..37B,2008ApJ...678..780G,2021MNRAS.502.4794N,2024MNRAS.532.4681P,2024ApJ...974..204K,2025MNRAS.537.3421R}. This introduces a source of scatter when relating stellar-scattering timescales to core size, and would introduce more variance into our model's $\log(M_{\rm def}/M_\star)$ distribution. Indeed, observations of the core galaxies A2261, IC 1101, NGC 4486B, and A0402 are consistent with recoiled SMBHs having enhanced the size of the core beyond initial binary scouring \citep{2012ApJ...756..159P, 2017MNRAS.471.2321D,2025arXiv251214695T,2026arXiv260310104M}.

Third, we ignore triple black hole interactions. Triplets are a natural consequence of hierarchical merging, especially if the coalescence timescale for SMBHBs is long compared to the galaxy merger timescale \citep{2018MNRAS.477.2599B}. Triplets are also a natural mechanism for prompt merging and often lead to tertiary ejection \citep{2025ApJ...993..222S}. These tertiary ejections have been shown to enhance the size of cores via post-kick oscillations in the nucleus \citep{2024MNRAS.527.7424S, 2025MNRAS.537.3421R}. Neglecting triples may therefore affect both the predicted binary occupation fraction and the mapping between core properties and hardening timescales.

Fourth, we do not consider how the sense of the black hole orbits relative to the stellar component influences hardening. Pairs on retrograde orbits may harden more rapidly than those on prograde orbits, lowering the expected mass deficit \citep{2015ApJ...810..139H, 2017ApJ...837..135R, 2017MNRAS.470..940M, 2021A&A...649A..41A,2025arXiv250814253H}. Including a distribution of orbital orientations could therefore produce additional single SMBH systems that overlap with those hosting binaries in the assumed feature space.

Fifth, our model does not include hot-gas dynamics. While the gas-poor nature of core ellipticals reduces the importance of cold circumbinary disks, massive ellipticals and BCGs often contain hot gaseous halos. Interactions with hot gas could affect orbital evolution or accretion signatures, which may be relevant when interpreting candidate systems observationally.

% Lastly, we draw attention to the fact that our sample of observed core galaxies is heterogeneous, composed of an over-representation of BCGs. This is largely reflected in the skew of the observed sample's $\log R_{\rm eff}/M_\star$ distribution compared to that of the model output toward higher values (see Figure \ref{fig:feature distributions}). This systematic shift in the feature space of the observed sample undoubtedly results in the model overestimating the binary host probability for BCGs. Indeed, our ensemble of logistic regression models insists A2052-BCG hosts a surviving SMBHB, likely owing to its $10^5 {\rm kpc}$ radius.

\section{Summary} 
\label{sec:summary}
We have used a core-scouring model applied to IllustrisTNG merger histories to estimate the probability that local core galaxies host supermassive black hole binaries. By combining simulated massive-galaxy assembly histories with an observationally motivated GSMF, we constructed a simulated population of core galaxies and performed logistic regression analyses to predict binary occupation from observable galaxy properties.

Our main conclusions are as follows:
\begin{enumerate}
    \item \emph{Core galaxies are rare but efficient binary hosts.} \\
    We find that the fraction of galaxies above $10^{10}M_\odot$ that host depleted cores is approximately 0.0019. Among core galaxies the predicted binary fraction is $0.18\pm0.02$.
    \item \emph{Binary occupation correlates with observable core galaxy properties.} \\
    Galaxies which are more diffuse, have lower stellar mass deficits for their stellar masses, and lower central densities for their velocity dispersions are more likely to host surviving binaries.
    \item \emph{A simple logistic regression model performs well.} \\
    Using three observable features our logistic regression models achieve a pseudo-$R^2$ value of $0.51 \pm0.03$ with accuracy $0.95 \pm 0.01$, precision $0.92 \pm 0.03$, and recall $0.82 \pm 0.04$ for a 50\% probability threshold.
    \item \emph{Surviving binaries are typically compact.} \\
    The predicted binary separation distribution peaks at $0.3 \ {\rm pc}$ with 16\% and 84\% confidence intervals of $0.1$ and $1 \ {\rm pc}$ respectively. Only 0.3\% of binaries across all model realizations emit in the PTA frequency band, however.
    \item \emph{Several observed core galaxies are promising binary candidates.} \\
    Applying our logistic regression models to the observed core galaxy population yields eight galaxies with estimated binary probabilities above 50\%. Five are distant brightest cluster galaxies. Messier 86 is the most promising nearby candidate.
\end{enumerate}
These results support the use of local core galaxies as a bridge between electromagnetic observations of galactic nuclei and the SMBHB population responsible for the nanohertz GWB. Future work incorporating eccentric orbits, triple interactions, GW-recoil, and gas physics will refine this connection and make target predictions more robust.
%% Please use the acknowledgment and contribution environments. This will 
%% be anonomyized when the "anonymous" style option is used. 
% \begin{acknowledgments}

% \end{acknowledgments}

\section{Data Availability}
The data underlying this work are deposited in Deep Blue Data, the University of Michigan's institutional data repository. The data will be publicly accessible and maintained for a minimum of ten years and can be accessed via https://doi.org/10.7302/zf4x-tv36.

\bibliography{references}{}

@ARTICLE{2024MNRAS.528....1H,
       author = {{Harris}, C.~J. and {G{\"u}ltekin}, Kayhan},
        title = "{Connecting core galaxy properties to the massive black hole binary population}",
      journal = {\mnras},
         year = 2024,
        month = feb,
       volume = {528},
       number = {1},
        pages = {1-10},
          doi = {10.1093/mnras/stad3589},
archivePrefix = {arXiv},
       eprint = {2311.04877},
 primaryClass = {astro-ph.GA},
       adsurl = {https://ui.adsabs.harvard.edu/abs/2024MNRAS.528....1H}
}

@ARTICLE{1980Natur.287..307B,
       author = {{Begelman}, M.~C. and {Blandford}, R.~D. and {Rees}, M.~J.},
        title = "{Massive black hole binaries in active galactic nuclei}",
      journal = {\nat},
         year = 1980,
        month = sep,
       volume = {287},
       number = {5780},
        pages = {307-309},
          doi = {10.1038/287307a0},
       adsurl = {https://ui.adsabs.harvard.edu/abs/1980Natur.287..307B}
}

@ARTICLE{2006ApJ...652..270B,
       author = {{Bell}, Eric F. and {Phleps}, Stefanie and {Somerville}, Rachel S. and {Wolf}, Christian and {Borch}, Andrea and {Meisenheimer}, Klaus},
        title = "{The Merger Rate of Massive Galaxies}",
      journal = {\apj},
         year = 2006,
        month = nov,
       volume = {652},
       number = {1},
        pages = {270-276},
          doi = {10.1086/508408},
archivePrefix = {arXiv},
       eprint = {astro-ph/0602038},
 primaryClass = {astro-ph},
       adsurl = {https://ui.adsabs.harvard.edu/abs/2006ApJ...652..270B}
}

@ARTICLE{2012A&A...548A...7L,
       author = {{L{\'o}pez-Sanjuan}, C. and {Le F{\`e}vre}, O. and {Ilbert}, O. and {Tasca}, L.~A.~M. and {Bridge}, C. and {Cucciati}, O. and {Kampczyk}, P. and {Pozzetti}, L. and {Xu}, C.~K. and {Carollo}, C.~M. and {Contini}, T. and {Kneib}, J. -P. and {Lilly}, S.~J. and {Mainieri}, V. and {Renzini}, A. and {Sanders}, D. and {Scodeggio}, M. and {Scoville}, N.~Z. and {Taniguchi}, Y. and {Zamorani}, G. and {Aussel}, H. and {Bardelli}, S. and {Bolzonella}, M. and {Bongiorno}, A. and {Capak}, P. and {Caputi}, K. and {de la Torre}, S. and {de Ravel}, L. and {Franzetti}, P. and {Garilli}, B. and {Iovino}, A. and {Knobel}, C. and {Kova{\v{c}}}, K. and {Lamareille}, F. and {Le Borgne}, J. -F. and {Le Brun}, V. and {Le Floc'h}, E. and {Maier}, C. and {McCracken}, H.~J. and {Mignoli}, M. and {Pell{\'o}}, R. and {Peng}, Y. and {P{\'e}rez-Montero}, E. and {Presotto}, V. and {Ricciardelli}, E. and {Salvato}, M. and {Silverman}, J.~D. and {Tanaka}, M. and {Tresse}, L. and {Vergani}, D. and {Zucca}, E. and {Barnes}, L. and {Bordoloi}, R. and {Cappi}, A. and {Cimatti}, A. and {Coppa}, G. and {Koekemoer}, A. and {Liu}, C.~T. and {Moresco}, M. and {Nair}, P. and {Oesch}, P. and {Schawinski}, K. and {Welikala}, N.},
        title = "{The dominant role of mergers in the size evolution of massive early-type galaxies since z \raisebox{-0.5ex}\textasciitilde 1}",
      journal = {\aap},
         year = 2012,
        month = dec,
       volume = {548},
          eid = {A7},
        pages = {A7},
          doi = {10.1051/0004-6361/201219085},
archivePrefix = {arXiv},
       eprint = {1202.4674},
 primaryClass = {astro-ph.CO},
       adsurl = {https://ui.adsabs.harvard.edu/abs/2012A&A...548A...7L}
}

@ARTICLE{2009ApJ...697.1369B,
       author = {{Bundy}, Kevin and {Fukugita}, Masataka and {Ellis}, Richard S. and {Targett}, Thomas A. and {Belli}, Sirio and {Kodama}, Tadayuki},
        title = "{The Greater Impact of Mergers on the Growth of Massive Galaxies: Implications for Mass Assembly and Evolution since z sime 1}",
      journal = {\apj},
         year = 2009,
        month = jun,
       volume = {697},
       number = {2},
        pages = {1369-1383},
          doi = {10.1088/0004-637X/697/2/1369},
archivePrefix = {arXiv},
       eprint = {0902.1188},
 primaryClass = {astro-ph.CO},
       adsurl = {https://ui.adsabs.harvard.edu/abs/2009ApJ...697.1369B}
}

@ARTICLE{2006ApJ...648..976M,
       author = {{Merritt}, David},
        title = "{Mass Deficits, Stalling Radii, and the Merger Histories of Elliptical Galaxies}",
      journal = {\apj},
         year = 2006,
        month = sep,
       volume = {648},
       number = {2},
        pages = {976-986},
          doi = {10.1086/506139},
archivePrefix = {arXiv},
       eprint = {astro-ph/0603439},
 primaryClass = {astro-ph},
       adsurl = {https://ui.adsabs.harvard.edu/abs/2006ApJ...648..976M}
}

@ARTICLE{2012ApJ...749..147K,
       author = {{Khan}, Fazeel Mahmood and {Preto}, Miguel and {Berczik}, Peter and {Berentzen}, Ingo and {Just}, Andreas and {Spurzem}, Rainer},
        title = "{Mergers of Unequal-mass Galaxies: Supermassive Black Hole Binary Evolution and Structure of Merger Remnants}",
      journal = {\apj},
         year = 2012,
        month = apr,
       volume = {749},
       number = {2},
          eid = {147},
        pages = {147},
          doi = {10.1088/0004-637X/749/2/147},
archivePrefix = {arXiv},
       eprint = {1202.2124},
 primaryClass = {astro-ph.CO},
       adsurl = {https://ui.adsabs.harvard.edu/abs/2012ApJ...749..147K}
}

@ARTICLE{2014ApJ...782...39T,
       author = {{Thomas}, J. and {Saglia}, R.~P. and {Bender}, R. and {Erwin}, P. and {Fabricius}, M.},
        title = "{The Dynamical Fingerprint of Core Scouring in Massive Elliptical Galaxies}",
      journal = {\apj},
         year = 2014,
        month = feb,
       volume = {782},
       number = {1},
          eid = {39},
        pages = {39},
          doi = {10.1088/0004-637X/782/1/39},
archivePrefix = {arXiv},
       eprint = {1311.3783},
 primaryClass = {astro-ph.GA},
       adsurl = {https://ui.adsabs.harvard.edu/abs/2014ApJ...782...39T}
}

@ARTICLE{2007ApJ...664..226L,
       author = {{Lauer}, Tod R. and {Gebhardt}, Karl and {Faber}, S.~M. and {Richstone}, Douglas and {Tremaine}, Scott and {Kormendy}, John and {Aller}, M.~C. and {Bender}, Ralf and {Dressler}, Alan and {Filippenko}, Alexei V. and {Green}, Richard and {Ho}, Luis C.},
        title = "{The Centers of Early-Type Galaxies with Hubble Space Telescope. VI. Bimodal Central Surface Brightness Profiles}",
      journal = {\apj},
         year = 2007,
        month = jul,
       volume = {664},
       number = {1},
        pages = {226-256},
          doi = {10.1086/519229},
archivePrefix = {arXiv},
       eprint = {astro-ph/0609762},
 primaryClass = {astro-ph},
       adsurl = {https://ui.adsabs.harvard.edu/abs/2007ApJ...664..226L}
}

@ARTICLE{2019ComAC...6....2N,
       author = {{Nelson}, Dylan and {Springel}, Volker and {Pillepich}, Annalisa and {Rodriguez-Gomez}, Vicente and {Torrey}, Paul and {Genel}, Shy and {Vogelsberger}, Mark and {Pakmor}, Ruediger and {Marinacci}, Federico and {Weinberger}, Rainer and {Kelley}, Luke and {Lovell}, Mark and {Diemer}, Benedikt and {Hernquist}, Lars},
        title = "{The IllustrisTNG simulations: public data release}",
      journal = {Computational Astrophysics and Cosmology},
         year = 2019,
        month = may,
       volume = {6},
       number = {1},
          eid = {2},
        pages = {2},
          doi = {10.1186/s40668-019-0028-x},
archivePrefix = {arXiv},
       eprint = {1812.05609},
 primaryClass = {astro-ph.GA},
       adsurl = {https://ui.adsabs.harvard.edu/abs/2019ComAC...6....2N}
}

@ARTICLE{1995AJ....110.2622L,
       author = {{Lauer}, T.~R. and {Ajhar}, E.~A. and {Byun}, Y. -I. and {Dressler}, A. and {Faber}, S.~M. and {Grillmair}, C. and {Kormendy}, J. and {Richstone}, D. and {Tremaine}, S.},
        title = "{The Centers of Early-Type Galaxies with HST.I.An Observational Survey}",
      journal = {\aj},
         year = 1995,
        month = dec,
       volume = {110},
        pages = {2622},
          doi = {10.1086/117719},
       adsurl = {https://ui.adsabs.harvard.edu/abs/1995AJ....110.2622L}
}

@ARTICLE{2017MNRAS.464.3131K,
       author = {{Kelley}, Luke Zoltan and {Blecha}, Laura and {Hernquist}, Lars},
        title = "{Massive black hole binary mergers in dynamical galactic environments}",
      journal = {\mnras},
         year = 2017,
        month = jan,
       volume = {464},
       number = {3},
        pages = {3131-3157},
          doi = {10.1093/mnras/stw2452},
archivePrefix = {arXiv},
       eprint = {1606.01900},
 primaryClass = {astro-ph.HE},
       adsurl = {https://ui.adsabs.harvard.edu/abs/2017MNRAS.464.3131K}
}

@ARTICLE{1999MNRAS.309..447M,
       author = {{Magorrian}, John and {Tremaine}, Scott},
        title = "{Rates of tidal disruption of stars by massive central black holes}",
      journal = {\mnras},
         year = 1999,
        month = oct,
       volume = {309},
       number = {2},
        pages = {447-460},
          doi = {10.1046/j.1365-8711.1999.02853.x},
archivePrefix = {arXiv},
       eprint = {astro-ph/9902032},
 primaryClass = {astro-ph},
       adsurl = {https://ui.adsabs.harvard.edu/abs/1999MNRAS.309..447M}
}

@ARTICLE{2013ARA&A..51..511K,
       author = {{Kormendy}, John and {Ho}, Luis C.},
        title = "{Coevolution (Or Not) of Supermassive Black Holes and Host Galaxies}",
      journal = {\araa},
         year = 2013,
        month = aug,
       volume = {51},
       number = {1},
        pages = {511-653},
          doi = {10.1146/annurev-astro-082708-101811},
archivePrefix = {arXiv},
       eprint = {1304.7762},
 primaryClass = {astro-ph.CO},
       adsurl = {https://ui.adsabs.harvard.edu/abs/2013ARA&A..51..511K}
}

@ARTICLE{2023ApJ...952L..37A,
       author = {{Agazie}, Gabriella and {Anumarlapudi}, Akash and {Archibald}, Anne M. and {Baker}, Paul T. and {B{\'e}csy}, Bence and {Blecha}, Laura and {Bonilla}, Alexander and {Brazier}, Adam and {Brook}, Paul R. and {Burke-Spolaor}, Sarah and {Burnette}, Rand and {Case}, Robin and {Casey-Clyde}, J. Andrew and {Charisi}, Maria and {Chatterjee}, Shami and {Chatziioannou}, Katerina and {Cheeseboro}, Belinda D. and {Chen}, Siyuan and {Cohen}, Tyler and {Cordes}, James M. and {Cornish}, Neil J. and {Crawford}, Fronefield and {Cromartie}, H. Thankful and {Crowter}, Kathryn and {Cutler}, Curt J. and {D'Orazio}, Daniel J. and {Decesar}, Megan E. and {Degan}, Dallas and {Demorest}, Paul B. and {Deng}, Heling and {Dolch}, Timothy and {Drachler}, Brendan and {Ferrara}, Elizabeth C. and {Fiore}, William and {Fonseca}, Emmanuel and {Freedman}, Gabriel E. and {Gardiner}, Emiko and {Garver-Daniels}, Nate and {Gentile}, Peter A. and {Gersbach}, Kyle A. and {Glaser}, Joseph and {Good}, Deborah C. and {G{\"u}ltekin}, Kayhan and {Hazboun}, Jeffrey S. and {Hourihane}, Sophie and {Islo}, Kristina and {Jennings}, Ross J. and {Johnson}, Aaron and {Jones}, Megan L. and {Kaiser}, Andrew R. and {Kaplan}, David L. and {Kelley}, Luke Zoltan and {Kerr}, Matthew and {Key}, Joey S. and {Laal}, Nima and {Lam}, Michael T. and {Lamb}, William G. and {Lazio}, T. Joseph W. and {Lewandowska}, Natalia and {Littenberg}, Tyson B. and {Liu}, Tingting and {Luo}, Jing and {Lynch}, Ryan S. and {Ma}, Chung-Pei and {Madison}, Dustin R. and {McEwen}, Alexander and {McKee}, James W. and {McLaughlin}, Maura A. and {McMann}, Natasha and {Meyers}, Bradley W. and {Meyers}, Patrick M. and {Mingarelli}, Chiara M.~F. and {Mitridate}, Andrea and {Natarajan}, Priyamvada and {Ng}, Cherry and {Nice}, David J. and {Ocker}, Stella Koch and {Olum}, Ken D. and {Pennucci}, Timothy T. and {Perera}, Benetge B.~P. and {Petrov}, Polina and {Pol}, Nihan S. and {Radovan}, Henri A. and {Ransom}, Scott M. and {Ray}, Paul S. and {Romano}, Joseph D. and {Runnoe}, Jessie C. and {Sardesai}, Shashwat C. and {Schmiedekamp}, Ann and {Schmiedekamp}, Carl and {Schmitz}, Kai and {Schult}, Levi and {Shapiro-Albert}, Brent J. and {Siemens}, Xavier and {Simon}, Joseph and {Siwek}, Magdalena S. and {Stairs}, Ingrid H. and {Stinebring}, Daniel R. and {Stovall}, Kevin and {Sun}, Jerry P. and {Susobhanan}, Abhimanyu and {Swiggum}, Joseph K. and {Taylor}, Jacob and {Taylor}, Stephen R. and {Turner}, Jacob E. and {Unal}, Caner and {Vallisneri}, Michele and {Vigeland}, Sarah J. and {Wachter}, Jeremy M. and {Wahl}, Haley M. and {Wang}, Qiaohong and {Witt}, Caitlin A. and {Wright}, David and {Young}, Olivia and {Nanograv Collaboration}},
        title = "{The NANOGrav 15 yr Data Set: Constraints on Supermassive Black Hole Binaries from the Gravitational-wave Background}",
      journal = {\apjl},
         year = 2023,
        month = aug,
       volume = {952},
       number = {2},
          eid = {L37},
        pages = {L37},
          doi = {10.3847/2041-8213/ace18b},
archivePrefix = {arXiv},
       eprint = {2306.16220},
 primaryClass = {astro-ph.HE},
       adsurl = {https://ui.adsabs.harvard.edu/abs/2023ApJ...952L..37A}
}

@ARTICLE{2023ApJ...951L...8A,
       author = {{Agazie}, Gabriella and {Anumarlapudi}, Akash and {Archibald}, Anne M. and {Arzoumanian}, Zaven and {Baker}, Paul T. and {B{\'e}csy}, Bence and {Blecha}, Laura and {Brazier}, Adam and {Brook}, Paul R. and {Burke-Spolaor}, Sarah and {Burnette}, Rand and {Case}, Robin and {Charisi}, Maria and {Chatterjee}, Shami and {Chatziioannou}, Katerina and {Cheeseboro}, Belinda D. and {Chen}, Siyuan and {Cohen}, Tyler and {Cordes}, James M. and {Cornish}, Neil J. and {Crawford}, Fronefield and {Cromartie}, H. Thankful and {Crowter}, Kathryn and {Cutler}, Curt J. and {Decesar}, Megan E. and {Degan}, Dallas and {Demorest}, Paul B. and {Deng}, Heling and {Dolch}, Timothy and {Drachler}, Brendan and {Ellis}, Justin A. and {Ferrara}, Elizabeth C. and {Fiore}, William and {Fonseca}, Emmanuel and {Freedman}, Gabriel E. and {Garver-Daniels}, Nate and {Gentile}, Peter A. and {Gersbach}, Kyle A. and {Glaser}, Joseph and {Good}, Deborah C. and {G{\"u}ltekin}, Kayhan and {Hazboun}, Jeffrey S. and {Hourihane}, Sophie and {Islo}, Kristina and {Jennings}, Ross J. and {Johnson}, Aaron D. and {Jones}, Megan L. and {Kaiser}, Andrew R. and {Kaplan}, David L. and {Kelley}, Luke Zoltan and {Kerr}, Matthew and {Key}, Joey S. and {Klein}, Tonia C. and {Laal}, Nima and {Lam}, Michael T. and {Lamb}, William G. and {Lazio}, T. Joseph W. and {Lewandowska}, Natalia and {Littenberg}, Tyson B. and {Liu}, Tingting and {Lommen}, Andrea and {Lorimer}, Duncan R. and {Luo}, Jing and {Lynch}, Ryan S. and {Ma}, Chung-Pei and {Madison}, Dustin R. and {Mattson}, Margaret A. and {McEwen}, Alexander and {McKee}, James W. and {McLaughlin}, Maura A. and {McMann}, Natasha and {Meyers}, Bradley W. and {Meyers}, Patrick M. and {Mingarelli}, Chiara M.~F. and {Mitridate}, Andrea and {Natarajan}, Priyamvada and {Ng}, Cherry and {Nice}, David J. and {Ocker}, Stella Koch and {Olum}, Ken D. and {Pennucci}, Timothy T. and {Perera}, Benetge B.~P. and {Petrov}, Polina and {Pol}, Nihan S. and {Radovan}, Henri A. and {Ransom}, Scott M. and {Ray}, Paul S. and {Romano}, Joseph D. and {Sardesai}, Shashwat C. and {Schmiedekamp}, Ann and {Schmiedekamp}, Carl and {Schmitz}, Kai and {Schult}, Levi and {Shapiro-Albert}, Brent J. and {Siemens}, Xavier and {Simon}, Joseph and {Siwek}, Magdalena S. and {Stairs}, Ingrid H. and {Stinebring}, Daniel R. and {Stovall}, Kevin and {Sun}, Jerry P. and {Susobhanan}, Abhimanyu and {Swiggum}, Joseph K. and {Taylor}, Jacob and {Taylor}, Stephen R. and {Turner}, Jacob E. and {Unal}, Caner and {Vallisneri}, Michele and {van Haasteren}, Rutger and {Vigeland}, Sarah J. and {Wahl}, Haley M. and {Wang}, Qiaohong and {Witt}, Caitlin A. and {Young}, Olivia and {Nanograv Collaboration}},
        title = "{The NANOGrav 15 yr Data Set: Evidence for a Gravitational-wave Background}",
      journal = {\apjl},
         year = 2023,
        month = jul,
       volume = {951},
       number = {1},
          eid = {L8},
        pages = {L8},
          doi = {10.3847/2041-8213/acdac6},
archivePrefix = {arXiv},
       eprint = {2306.16213},
 primaryClass = {astro-ph.HE},
       adsurl = {https://ui.adsabs.harvard.edu/abs/2023ApJ...951L...8A}
}

@ARTICLE{2006astro.ph..1520H,
       author = {{Holley-Bockelmann}, Kelly and {Sigurdsson}, Steinn},
        title = "{A Full Loss Cone For Triaxial Galaxies}",
      journal = {arXiv e-prints},
         year = 2006,
        month = jan,
          eid = {astro-ph/0601520},
        pages = {astro-ph/0601520},
          doi = {10.48550/arXiv.astro-ph/0601520},
archivePrefix = {arXiv},
       eprint = {astro-ph/0601520},
 primaryClass = {astro-ph},
       adsurl = {https://ui.adsabs.harvard.edu/abs/2006astro.ph..1520H}
}

@ARTICLE{2015ApJ...810..139H,
       author = {{Holley-Bockelmann}, Kelly and {Khan}, Fazeel Mahmood},
        title = "{Galaxy Rotation and Rapid Supermassive Binary Coalescence}",
      journal = {\apj},
         year = 2015,
        month = sep,
       volume = {810},
       number = {2},
          eid = {139},
        pages = {139},
          doi = {10.1088/0004-637X/810/2/139},
archivePrefix = {arXiv},
       eprint = {1505.06203},
 primaryClass = {astro-ph.GA},
       adsurl = {https://ui.adsabs.harvard.edu/abs/2015ApJ...810..139H}
}

@ARTICLE{2005MNRAS.362..184B,
       author = {{Boylan-Kolchin}, Michael and {Ma}, Chung-Pei and {Quataert}, Eliot},
        title = "{Dissipationless mergers of elliptical galaxies and the evolution of the fundamental plane}",
      journal = {\mnras},
         year = 2005,
        month = sep,
       volume = {362},
       number = {1},
        pages = {184-196},
          doi = {10.1111/j.1365-2966.2005.09278.x},
archivePrefix = {arXiv},
       eprint = {astro-ph/0502495},
 primaryClass = {astro-ph},
       adsurl = {https://ui.adsabs.harvard.edu/abs/2005MNRAS.362..184B}
}

@ARTICLE{2009ApJS..181..486H,
       author = {{Hopkins}, Philip F. and {Lauer}, Tod R. and {Cox}, Thomas J. and {Hernquist}, Lars and {Kormendy}, John},
        title = "{Dissipation and Extra Light in Galactic Nuclei. III. ``Core'' Ellipticals and ``Missing'' Light}",
      journal = {\apjs},
         year = 2009,
        month = apr,
       volume = {181},
       number = {2},
        pages = {486-532},
          doi = {10.1088/0067-0049/181/2/486},
archivePrefix = {arXiv},
       eprint = {0806.2325},
 primaryClass = {astro-ph},
       adsurl = {https://ui.adsabs.harvard.edu/abs/2009ApJS..181..486H}
}

@ARTICLE{2007ApJ...662..808L,
       author = {{Lauer}, Tod R. and {Faber}, S.~M. and {Richstone}, Douglas and {Gebhardt}, Karl and {Tremaine}, Scott and {Postman}, Marc and {Dressler}, Alan and {Aller}, M.~C. and {Filippenko}, Alexei V. and {Green}, Richard and {Ho}, Luis C. and {Kormendy}, John and {Magorrian}, John and {Pinkney}, Jason},
        title = "{The Masses of Nuclear Black Holes in Luminous Elliptical Galaxies and Implications for the Space Density of the Most Massive Black Holes}",
      journal = {\apj},
         year = 2007,
        month = jun,
       volume = {662},
       number = {2},
        pages = {808-834},
          doi = {10.1086/518223},
archivePrefix = {arXiv},
       eprint = {astro-ph/0606739},
 primaryClass = {astro-ph},
       adsurl = {https://ui.adsabs.harvard.edu/abs/2007ApJ...662..808L}
}

@ARTICLE{2021ApJ...922...40D,
       author = {{Dosopoulou}, Fani and {Greene}, Jenny E. and {Ma}, Chung-Pei},
        title = "{Galaxy Core Formation by Supermassive Black Hole Binaries: The Importance of Realistic Initial Conditions and Galaxy Morphology}",
      journal = {\apj},
         year = 2021,
        month = nov,
       volume = {922},
       number = {1},
          eid = {40},
        pages = {40},
          doi = {10.3847/1538-4357/ac1fe4},
archivePrefix = {arXiv},
       eprint = {2108.08317},
 primaryClass = {astro-ph.GA},
       adsurl = {https://ui.adsabs.harvard.edu/abs/2021ApJ...922...40D}
}

@ARTICLE{2017MNRAS.464.2301G,
       author = {{Gualandris}, Alessia and {Read}, Justin I. and {Dehnen}, Walter and {Bortolas}, Elisa},
        title = "{Collisionless loss-cone refilling: there is no final parsec problem}",
      journal = {\mnras},
         year = 2017,
        month = jan,
       volume = {464},
       number = {2},
        pages = {2301-2310},
          doi = {10.1093/mnras/stw2528},
archivePrefix = {arXiv},
       eprint = {1609.09383},
 primaryClass = {astro-ph.GA},
       adsurl = {https://ui.adsabs.harvard.edu/abs/2017MNRAS.464.2301G}
}

@ARTICLE{2006ApJ...651..392S,
       author = {{Sesana}, Alberto and {Haardt}, Francesco and {Madau}, Piero},
        title = "{Interaction of Massive Black Hole Binaries with Their Stellar Environment. I. Ejection of Hypervelocity Stars}",
      journal = {\apj},
         year = 2006,
        month = nov,
       volume = {651},
       number = {1},
        pages = {392-400},
          doi = {10.1086/507596},
archivePrefix = {arXiv},
       eprint = {astro-ph/0604299},
 primaryClass = {astro-ph},
       adsurl = {https://ui.adsabs.harvard.edu/abs/2006ApJ...651..392S}
}

@ARTICLE{1996NewA....1...35Q,
       author = {{Quinlan}, Gerald D.},
        title = "{The dynamical evolution of massive black hole binaries I. Hardening in a fixed stellar background}",
      journal = {\na},
         year = 1996,
        month = jul,
       volume = {1},
       number = {1},
        pages = {35-56},
          doi = {10.1016/S1384-1076(96)00003-6},
archivePrefix = {arXiv},
       eprint = {astro-ph/9601092},
 primaryClass = {astro-ph},
       adsurl = {https://ui.adsabs.harvard.edu/abs/1996NewA....1...35Q}
}

@ARTICLE{2014ApJ...788...11L,
       author = {{Lang}, Philipp and {Wuyts}, Stijn and {Somerville}, Rachel S. and {F{\"o}rster Schreiber}, Natascha M. and {Genzel}, Reinhard and {Bell}, Eric F. and {Brammer}, Gabe and {Dekel}, Avishai and {Faber}, Sandra M. and {Ferguson}, Henry C. and {Grogin}, Norman A. and {Kocevski}, Dale D. and {Koekemoer}, Anton M. and {Lutz}, Dieter and {McGrath}, Elizabeth J. and {Momcheva}, Ivelina and {Nelson}, Erica J. and {Primack}, Joel R. and {Rosario}, David J. and {Skelton}, Rosalind E. and {Tacconi}, Linda J. and {van Dokkum}, Pieter G. and {Whitaker}, Katherine E.},
        title = "{Bulge Growth and Quenching since z = 2.5 in CANDELS/3D-HST}",
      journal = {\apj},
         year = 2014,
        month = jun,
       volume = {788},
       number = {1},
          eid = {11},
        pages = {11},
          doi = {10.1088/0004-637X/788/1/11},
archivePrefix = {arXiv},
       eprint = {1402.0866},
 primaryClass = {astro-ph.GA},
       adsurl = {https://ui.adsabs.harvard.edu/abs/2014ApJ...788...11L}
}

@ARTICLE{2014MNRAS.441..599B,
       author = {{Bluck}, Asa F.~L. and {Mendel}, J. Trevor and {Ellison}, Sara L. and {Moreno}, Jorge and {Simard}, Luc and {Patton}, David R. and {Starkenburg}, Else},
        title = "{Bulge mass is king: the dominant role of the bulge in determining the fraction of passive galaxies in the Sloan Digital Sky Survey}",
      journal = {\mnras},
         year = 2014,
        month = jun,
       volume = {441},
       number = {1},
        pages = {599-629},
          doi = {10.1093/mnras/stu594},
archivePrefix = {arXiv},
       eprint = {1403.5269},
 primaryClass = {astro-ph.GA},
       adsurl = {https://ui.adsabs.harvard.edu/abs/2014MNRAS.441..599B}
}

@ARTICLE{2019ApJ...875L...1E,
       author = {{Event Horizon Telescope Collaboration} and {Akiyama}, Kazunori and {Alberdi}, Antxon and {Alef}, Walter and {Asada}, Keiichi and {Azulay}, Rebecca and {Baczko}, Anne-Kathrin and {Ball}, David and {Balokovi{\'c}}, Mislav and {Barrett}, John and {Bintley}, Dan and {Blackburn}, Lindy and {Boland}, Wilfred and {Bouman}, Katherine L. and {Bower}, Geoffrey C. and {Bremer}, Michael and {Brinkerink}, Christiaan D. and {Brissenden}, Roger and {Britzen}, Silke and {Broderick}, Avery E. and {Broguiere}, Dominique and {Bronzwaer}, Thomas and {Byun}, Do-Young and {Carlstrom}, John E. and {Chael}, Andrew and {Chan}, Chi-kwan and {Chatterjee}, Shami and {Chatterjee}, Koushik and {Chen}, Ming-Tang and {Chen}, Yongjun and {Cho}, Ilje and {Christian}, Pierre and {Conway}, John E. and {Cordes}, James M. and {Crew}, Geoffrey B. and {Cui}, Yuzhu and {Davelaar}, Jordy and {De Laurentis}, Mariafelicia and {Deane}, Roger and {Dempsey}, Jessica and {Desvignes}, Gregory and {Dexter}, Jason and {Doeleman}, Sheperd S. and {Eatough}, Ralph P. and {Falcke}, Heino and {Fish}, Vincent L. and {Fomalont}, Ed and {Fraga-Encinas}, Raquel and {Freeman}, William T. and {Friberg}, Per and {Fromm}, Christian M. and {G{\'o}mez}, Jos{\'e} L. and {Galison}, Peter and {Gammie}, Charles F. and {Garc{\'\i}a}, Roberto and {Gentaz}, Olivier and {Georgiev}, Boris and {Goddi}, Ciriaco and {Gold}, Roman and {Gu}, Minfeng and {Gurwell}, Mark and {Hada}, Kazuhiro and {Hecht}, Michael H. and {Hesper}, Ronald and {Ho}, Luis C. and {Ho}, Paul and {Honma}, Mareki and {Huang}, Chih-Wei L. and {Huang}, Lei and {Hughes}, David H. and {Ikeda}, Shiro and {Inoue}, Makoto and {Issaoun}, Sara and {James}, David J. and {Jannuzi}, Buell T. and {Janssen}, Michael and {Jeter}, Britton and {Jiang}, Wu and {Johnson}, Michael D. and {Jorstad}, Svetlana and {Jung}, Taehyun and {Karami}, Mansour and {Karuppusamy}, Ramesh and {Kawashima}, Tomohisa and {Keating}, Garrett K. and {Kettenis}, Mark and {Kim}, Jae-Young and {Kim}, Junhan and {Kim}, Jongsoo and {Kino}, Motoki and {Koay}, Jun Yi and {Koch}, Patrick M. and {Koyama}, Shoko and {Kramer}, Michael and {Kramer}, Carsten and {Krichbaum}, Thomas P. and {Kuo}, Cheng-Yu and {Lauer}, Tod R. and {Lee}, Sang-Sung and {Li}, Yan-Rong and {Li}, Zhiyuan and {Lindqvist}, Michael and {Liu}, Kuo and {Liuzzo}, Elisabetta and {Lo}, Wen-Ping and {Lobanov}, Andrei P. and {Loinard}, Laurent and {Lonsdale}, Colin and {Lu}, Ru-Sen and {MacDonald}, Nicholas R. and {Mao}, Jirong and {Markoff}, Sera and {Marrone}, Daniel P. and {Marscher}, Alan P. and {Mart{\'\i}-Vidal}, Iv{\'a}n and {Matsushita}, Satoki and {Matthews}, Lynn D. and {Medeiros}, Lia and {Menten}, Karl M. and {Mizuno}, Yosuke and {Mizuno}, Izumi and {Moran}, James M. and {Moriyama}, Kotaro and {Moscibrodzka}, Monika and {M{\"u}ller}, Cornelia and {Nagai}, Hiroshi and {Nagar}, Neil M. and {Nakamura}, Masanori and {Narayan}, Ramesh and {Narayanan}, Gopal and {Natarajan}, Iniyan and {Neri}, Roberto and {Ni}, Chunchong and {Noutsos}, Aristeidis and {Okino}, Hiroki and {Olivares}, H{\'e}ctor and {Ortiz-Le{\'o}n}, Gisela N. and {Oyama}, Tomoaki and {{\"O}zel}, Feryal and {Palumbo}, Daniel C.~M. and {Patel}, Nimesh and {Pen}, Ue-Li and {Pesce}, Dominic W. and {Pi{\'e}tu}, Vincent and {Plambeck}, Richard and {PopStefanija}, Aleksandar and {Porth}, Oliver and {Prather}, Ben and {Preciado-L{\'o}pez}, Jorge A. and {Psaltis}, Dimitrios and {Pu}, Hung-Yi and {Ramakrishnan}, Venkatessh and {Rao}, Ramprasad and {Rawlings}, Mark G. and {Raymond}, Alexander W. and {Rezzolla}, Luciano and {Ripperda}, Bart and {Roelofs}, Freek and {Rogers}, Alan and {Ros}, Eduardo and {Rose}, Mel and {Roshanineshat}, Arash and {Rottmann}, Helge and {Roy}, Alan L. and {Ruszczyk}, Chet and {Ryan}, Benjamin R. and {Rygl}, Kazi L.~J. and {S{\'a}nchez}, Salvador and {S{\'a}nchez-Arguelles}, David and {Sasada}, Mahito and {Savolainen}, Tuomas and {Schloerb}, F. Peter and {Schuster}, Karl-Friedrich and {Shao}, Lijing and {Shen}, Zhiqiang and {Small}, Des and {Sohn}, Bong Won and {SooHoo}, Jason and {Tazaki}, Fumie and {Tiede}, Paul and {Tilanus}, Remo P.~J. and {Titus}, Michael and {Toma}, Kenji and {Torne}, Pablo and {Trent}, Tyler and {Trippe}, Sascha and {Tsuda}, Shuichiro and {van Bemmel}, Ilse and {van Langevelde}, Huib Jan and {van Rossum}, Daniel R. and {Wagner}, Jan and {Wardle}, John and {Weintroub}, Jonathan and {Wex}, Norbert and {Wharton}, Robert and {Wielgus}, Maciek and {Wong}, George N. and {Wu}, Qingwen and {Young}, Ken and {Young}, Andr{\'e}},
        title = "{First M87 Event Horizon Telescope Results. I. The Shadow of the Supermassive Black Hole}",
      journal = {\apjl},
         year = 2019,
        month = apr,
       volume = {875},
       number = {1},
          eid = {L1},
        pages = {L1},
          doi = {10.3847/2041-8213/ab0ec7},
archivePrefix = {arXiv},
       eprint = {1906.11238},
 primaryClass = {astro-ph.GA},
       adsurl = {https://ui.adsabs.harvard.edu/abs/2019ApJ...875L...1E}
}

@ARTICLE{2008MNRAS.386..864D,
       author = {{Dabringhausen}, J. and {Hilker}, M. and {Kroupa}, P.},
        title = "{From star clusters to dwarf galaxies: the properties of dynamically hot stellar systems}",
      journal = {\mnras},
         year = 2008,
        month = may,
       volume = {386},
       number = {2},
        pages = {864-886},
          doi = {10.1111/j.1365-2966.2008.13065.x},
archivePrefix = {arXiv},
       eprint = {0802.0703},
 primaryClass = {astro-ph},
       adsurl = {https://ui.adsabs.harvard.edu/abs/2008MNRAS.386..864D}
}

@ARTICLE{1993MNRAS.265..250D,
       author = {{Dehnen}, W.},
        title = "{A Family of Potential-Density Pairs for Spherical Galaxies and Bulges}",
      journal = {\mnras},
         year = 1993,
        month = nov,
       volume = {265},
        pages = {250},
          doi = {10.1093/mnras/265.1.250},
       adsurl = {https://ui.adsabs.harvard.edu/abs/1993MNRAS.265..250D}
}

@ARTICLE{2009ApJS..181..135H,
       author = {{Hopkins}, Philip F. and {Cox}, Thomas J. and {Dutta}, Suvendra N. and {Hernquist}, Lars and {Kormendy}, John and {Lauer}, Tod R.},
        title = "{Dissipation and Extra Light in Galactic Nuclei. II. ``Cusp'' Ellipticals}",
      journal = {\apjs},
         year = 2009,
        month = mar,
       volume = {181},
       number = {1},
        pages = {135-182},
          doi = {10.1088/0067-0049/181/1/135},
archivePrefix = {arXiv},
       eprint = {0805.3533},
 primaryClass = {astro-ph},
       adsurl = {https://ui.adsabs.harvard.edu/abs/2009ApJS..181..135H}
}

@ARTICLE{2016MNRAS.457.4340K,
       author = {{Klypin}, Anatoly and {Yepes}, Gustavo and {Gottl{\"o}ber}, Stefan and {Prada}, Francisco and {He{\ss}}, Steffen},
        title = "{MultiDark simulations: the story of dark matter halo concentrations and density profiles}",
      journal = {\mnras},
         year = 2016,
        month = apr,
       volume = {457},
       number = {4},
        pages = {4340-4359},
          doi = {10.1093/mnras/stw248},
archivePrefix = {arXiv},
       eprint = {1411.4001},
 primaryClass = {astro-ph.CO},
       adsurl = {https://ui.adsabs.harvard.edu/abs/2016MNRAS.457.4340K}
}

@ARTICLE{2005AJ....129.2138L,
       author = {{Lauer}, Tod R. and {Faber}, S.~M. and {Gebhardt}, Karl and {Richstone}, Douglas and {Tremaine}, Scott and {Ajhar}, Edward A. and {Aller}, M.~C. and {Bender}, Ralf and {Dressler}, Alan and {Filippenko}, Alexei V. and {Green}, Richard and {Grillmair}, Carl J. and {Ho}, Luis C. and {Kormendy}, John and {Magorrian}, John and {Pinkney}, Jason and {Siopis}, Christos},
        title = "{The Centers of Early-Type Galaxies with Hubble Space Telescope. V. New WFPC2 Photometry}",
      journal = {\aj},
         year = 2005,
        month = may,
       volume = {129},
       number = {5},
        pages = {2138-2185},
          doi = {10.1086/429565},
archivePrefix = {arXiv},
       eprint = {astro-ph/0412040},
 primaryClass = {astro-ph},
       adsurl = {https://ui.adsabs.harvard.edu/abs/2005AJ....129.2138L}
}

@ARTICLE{2001AJ....121.2431R,
       author = {{Rest}, Armin and {van den Bosch}, Frank C. and {Jaffe}, Walter and {Tran}, Hien and {Tsvetanov}, Zlatan and {Ford}, Holland C. and {Davies}, James and {Schafer}, Joanna},
        title = "{WFPC2 Images of the Central Regions of Early-Type Galaxies. I. The Data}",
      journal = {\aj},
         year = 2001,
        month = may,
       volume = {121},
       number = {5},
        pages = {2431-2482},
          doi = {10.1086/320370},
archivePrefix = {arXiv},
       eprint = {astro-ph/0102286},
 primaryClass = {astro-ph},
       adsurl = {https://ui.adsabs.harvard.edu/abs/2001AJ....121.2431R}
}

@ARTICLE{2003AJ....125..478L,
       author = {{Laine}, Seppo and {van der Marel}, Roeland P. and {Lauer}, Tod R. and {Postman}, Marc and {O'Dea}, Christopher P. and {Owen}, Frazer N.},
        title = "{Hubble Space Telescope Imaging of Brightest Cluster Galaxies}",
      journal = {\aj},
         year = 2003,
        month = feb,
       volume = {125},
       number = {2},
        pages = {478-505},
          doi = {10.1086/345823},
archivePrefix = {arXiv},
       eprint = {astro-ph/0211074},
 primaryClass = {astro-ph},
       adsurl = {https://ui.adsabs.harvard.edu/abs/2003AJ....125..478L}
}

@ARTICLE{1997AJ....114.1771F,
       author = {{Faber}, S.~M. and {Tremaine}, Scott and {Ajhar}, Edward A. and {Byun}, Yong-Ik and {Dressler}, Alan and {Gebhardt}, Karl and {Grillmair}, Carl and {Kormendy}, John and {Lauer}, Tod R. and {Richstone}, Douglas},
        title = "{The Centers of Early-Type Galaxies with HST. IV. Central Parameter Relations.}",
      journal = {\aj},
         year = 1997,
        month = nov,
       volume = {114},
        pages = {1771},
          doi = {10.1086/118606},
archivePrefix = {arXiv},
       eprint = {astro-ph/9610055},
 primaryClass = {astro-ph},
       adsurl = {https://ui.adsabs.harvard.edu/abs/1997AJ....114.1771F}
}

@ARTICLE{2001AJ....122..653R,
       author = {{Ravindranath}, Swara and {Ho}, Luis C. and {Peng}, Chien Y. and {Filippenko}, Alexei V. and {Sargent}, Wallace L.~W.},
        title = "{Central Structural Parameters of Early-Type Galaxies as Viewed with Nicmos on the Hubble Space Telescope}",
      journal = {\aj},
         year = 2001,
        month = aug,
       volume = {122},
       number = {2},
        pages = {653-678},
          doi = {10.1086/321175},
archivePrefix = {arXiv},
       eprint = {astro-ph/0105390},
 primaryClass = {astro-ph},
       adsurl = {https://ui.adsabs.harvard.edu/abs/2001AJ....122..653R}
}

@ARTICLE{2000ApJS..128...85Q,
       author = {{Quillen}, A.~C. and {Bower}, Gary A. and {Stritzinger}, M.},
        title = "{A NICMOS Survey of Early-Type Galaxy Centers: The Relation Between Core Properties, Gas and Dust Content, and Environment}",
      journal = {\apjs},
         year = 2000,
        month = may,
       volume = {128},
       number = {1},
        pages = {85-98},
          doi = {10.1086/313374},
archivePrefix = {arXiv},
       eprint = {astro-ph/9907021},
 primaryClass = {astro-ph},
       adsurl = {https://ui.adsabs.harvard.edu/abs/2000ApJS..128...85Q}
}

@ARTICLE{1998AJ....115.2285M,
       author = {{Magorrian}, John and {Tremaine}, Scott and {Richstone}, Douglas and {Bender}, Ralf and {Bower}, Gary and {Dressler}, Alan and {Faber}, S.~M. and {Gebhardt}, Karl and {Green}, Richard and {Grillmair}, Carl and {Kormendy}, John and {Lauer}, Tod},
        title = "{The Demography of Massive Dark Objects in Galaxy Centers}",
      journal = {\aj},
         year = 1998,
        month = jun,
       volume = {115},
       number = {6},
        pages = {2285-2305},
          doi = {10.1086/300353},
archivePrefix = {arXiv},
       eprint = {astro-ph/9708072},
 primaryClass = {astro-ph},
       adsurl = {https://ui.adsabs.harvard.edu/abs/1998AJ....115.2285M}
}

@ARTICLE{2003AJ....125.2951G,
       author = {{Graham}, Alister W. and {Erwin}, Peter and {Trujillo}, I. and {Asensio Ramos}, A.},
        title = "{A New Empirical Model for the Structural Analysis of Early-Type Galaxies, and A Critical Review of the Nuker Model}",
      journal = {\aj},
         year = 2003,
        month = jun,
       volume = {125},
       number = {6},
        pages = {2951-2963},
          doi = {10.1086/375320},
archivePrefix = {arXiv},
       eprint = {astro-ph/0306023},
 primaryClass = {astro-ph},
       adsurl = {https://ui.adsabs.harvard.edu/abs/2003AJ....125.2951G}
}

@ARTICLE{2006ApJS..164..334F,
       author = {{Ferrarese}, Laura and {C{\^o}t{\'e}}, Patrick and {Jord{\'a}n}, Andr{\'e}s and {Peng}, Eric W. and {Blakeslee}, John P. and {Piatek}, Slawomir and {Mei}, Simona and {Merritt}, David and {Milosavljevi{\'c}}, Milo{\v{s}} and {Tonry}, John L. and {West}, Michael J.},
        title = "{The ACS Virgo Cluster Survey. VI. Isophotal Analysis and the Structure of Early-Type Galaxies}",
      journal = {\apjs},
         year = 2006,
        month = jun,
       volume = {164},
       number = {2},
        pages = {334-434},
          doi = {10.1086/501350},
archivePrefix = {arXiv},
       eprint = {astro-ph/0602297},
 primaryClass = {astro-ph},
       adsurl = {https://ui.adsabs.harvard.edu/abs/2006ApJS..164..334F}
}

@ARTICLE{1998Natur.395A..14R,
       author = {{Richstone}, D. and {Ajhar}, E.~A. and {Bender}, R. and {Bower}, G. and {Dressler}, A. and {Faber}, S.~M. and {Filippenko}, A.~V. and {Gebhardt}, K. and {Green}, R. and {Ho}, L.~C. and {Kormendy}, J. and {Lauer}, T.~R. and {Magorrian}, J. and {Tremaine}, S.},
        title = "{Supermassive black holes and the evolution of galaxies.}",
      journal = {\nat},
         year = 1998,
        month = oct,
       volume = {385},
       number = {6701},
        pages = {A14},
          doi = {10.48550/arXiv.astro-ph/9810378},
archivePrefix = {arXiv},
       eprint = {astro-ph/9810378},
 primaryClass = {astro-ph},
       adsurl = {https://ui.adsabs.harvard.edu/abs/1998Natur.395A..14R}
}

@ARTICLE{2011MNRAS.411.1467K,
       author = {{Kocsis}, B. and {Sesana}, A.},
        title = "{Gas-driven massive black hole binaries: signatures in the nHz gravitational wave background}",
      journal = {\mnras},
         year = 2011,
        month = mar,
       volume = {411},
       number = {3},
        pages = {1467-1479},
          doi = {10.1111/j.1365-2966.2010.17782.x},
archivePrefix = {arXiv},
       eprint = {1002.0584},
 primaryClass = {astro-ph.CO},
       adsurl = {https://ui.adsabs.harvard.edu/abs/2011MNRAS.411.1467K}
}

@ARTICLE{2012AJ....143..119I,
       author = {{Inada}, Naohisa and {Oguri}, Masamune and {Shin}, Min-Su and {Kayo}, Issha and {Strauss}, Michael A. and {Morokuma}, Tomoki and {Rusu}, Cristian E. and {Fukugita}, Masataka and {Kochanek}, Christopher S. and {Richards}, Gordon T. and {Schneider}, Donald P. and {York}, Donald G. and {Bahcall}, Neta A. and {Frieman}, Joshua A. and {Hall}, Patrick B. and {White}, Richard L.},
        title = "{The Sloan Digital Sky Survey Quasar Lens Search. V. Final Catalog from the Seventh Data Release}",
      journal = {\aj},
         year = 2012,
        month = may,
       volume = {143},
       number = {5},
          eid = {119},
        pages = {119},
          doi = {10.1088/0004-6256/143/5/119},
archivePrefix = {arXiv},
       eprint = {1203.1087},
 primaryClass = {astro-ph.CO},
       adsurl = {https://ui.adsabs.harvard.edu/abs/2012AJ....143..119I}
}

@ARTICLE{2016MNRAS.456.1595M,
       author = {{More}, Anupreeta and {Oguri}, Masamune and {Kayo}, Issha and {Zinn}, Joel and {Strauss}, Michael A. and {Santiago}, Basilio X. and {Mosquera}, Ana M. and {Inada}, Naohisa and {Kochanek}, Christopher S. and {Rusu}, Cristian E. and {Brownstein}, Joel R. and {da Costa}, Luiz N. and {Kneib}, Jean-Paul and {Maia}, Marcio A.~G. and {Quimby}, Robert M. and {Schneider}, Donald P. and {Streblyanska}, Alina and {York}, Donald G.},
        title = "{The SDSS-III BOSS quasar lens survey: discovery of 13 gravitationally lensed quasars}",
      journal = {\mnras},
         year = 2016,
        month = feb,
       volume = {456},
       number = {2},
        pages = {1595-1606},
          doi = {10.1093/mnras/stv2813},
archivePrefix = {arXiv},
       eprint = {1509.07917},
 primaryClass = {astro-ph.GA},
       adsurl = {https://ui.adsabs.harvard.edu/abs/2016MNRAS.456.1595M}
}

@ARTICLE{2019MNRAS.483.4242L,
       author = {{Lemon}, Cameron A. and {Auger}, Matthew W. and {McMahon}, Richard G.},
        title = "{Gravitationally lensed quasars in Gaia - III. 22 new lensed quasars from Gaia data release 2}",
      journal = {\mnras},
         year = 2019,
        month = mar,
       volume = {483},
       number = {3},
        pages = {4242-4258},
          doi = {10.1093/mnras/sty3366},
archivePrefix = {arXiv},
       eprint = {1810.04480},
 primaryClass = {astro-ph.GA},
       adsurl = {https://ui.adsabs.harvard.edu/abs/2019MNRAS.483.4242L}
}

@ARTICLE{2019NewAR..8601525D,
       author = {{De Rosa}, Alessandra and {Vignali}, Cristian and {Bogdanovi{\'c}}, Tamara and {Capelo}, Pedro R. and {Charisi}, Maria and {Dotti}, Massimo and {Husemann}, Bernd and {Lusso}, Elisabeta and {Mayer}, Lucio and {Paragi}, Zsolt and {Runnoe}, Jessie and {Sesana}, Alberto and {Steinborn}, Lisa and {Bianchi}, Stefano and {Colpi}, Monica and {del Valle}, Luciano and {Frey}, S{\'a}ndor and {Gab{\'a}nyi}, Krisztina {\'E}. and {Giustini}, Margherita and {Guainazzi}, Matteo and {Haiman}, Zoltan and {Herrera Ruiz}, Noelia and {Herrero-Illana}, Rub{\'e}n and {Iwasawa}, Kazushi and {Komossa}, S. and {Lena}, Davide and {Loiseau}, Nora and {Perez-Torres}, Miguel and {Piconcelli}, Enrico and {Volonteri}, Marta},
        title = "{The quest for dual and binary supermassive black holes: A multi-messenger view}",
      journal = {\nar},
         year = 2019,
        month = dec,
       volume = {86},
          eid = {101525},
        pages = {101525},
          doi = {10.1016/j.newar.2020.101525},
archivePrefix = {arXiv},
       eprint = {2001.06293},
 primaryClass = {astro-ph.GA},
       adsurl = {https://ui.adsabs.harvard.edu/abs/2019NewAR..8601525D}
}

@ARTICLE{2005SSRv..116..523F,
       author = {{Ferrarese}, Laura and {Ford}, Holland},
        title = "{Supermassive Black Holes in Galactic Nuclei: Past, Present and Future Research}",
      journal = {\ssr},
         year = 2005,
        month = feb,
       volume = {116},
       number = {3-4},
        pages = {523-624},
          doi = {10.1007/s11214-005-3947-6},
archivePrefix = {arXiv},
       eprint = {astro-ph/0411247},
 primaryClass = {astro-ph},
       adsurl = {https://ui.adsabs.harvard.edu/abs/2005SSRv..116..523F}
}

@ARTICLE{2019ApJ...877...17F,
       author = {{Foord}, Adi and {G{\"u}ltekin}, Kayhan and {Reynolds}, Mark T. and {Hodges-Kluck}, Edmund and {Cackett}, Edward M. and {Comerford}, Julia M. and {King}, Ashley L. and {Miller}, Jon M. and {Runnoe}, Jessie C.},
        title = "{A Bayesian Analysis of SDSS J0914+0853, a Low-mass Dual AGN Candidate}",
      journal = {\apj},
         year = 2019,
        month = may,
       volume = {877},
       number = {1},
          eid = {17},
        pages = {17},
          doi = {10.3847/1538-4357/ab18a3},
archivePrefix = {arXiv},
       eprint = {1904.06363},
 primaryClass = {astro-ph.HE},
       adsurl = {https://ui.adsabs.harvard.edu/abs/2019ApJ...877...17F}
}

@ARTICLE{2020ApJ...892...29F,
       author = {{Foord}, Adi and {G{\"u}ltekin}, Kayhan and {Nevin}, Rebecca and {Comerford}, Julia M. and {Hodges-Kluck}, Edmund and {Barrows}, R. Scott and {Goulding}, Andrew D. and {Greene}, Jenny E.},
        title = "{A Second Look at 12 Candidate Dual AGNs Using BAYMAX}",
      journal = {\apj},
         year = 2020,
        month = mar,
       volume = {892},
       number = {1},
          eid = {29},
        pages = {29},
          doi = {10.3847/1538-4357/ab72fa},
archivePrefix = {arXiv},
       eprint = {2002.01033},
 primaryClass = {astro-ph.HE},
       adsurl = {https://ui.adsabs.harvard.edu/abs/2020ApJ...892...29F}
}

@ARTICLE{2021ApJ...907...71F,
       author = {{Foord}, Adi and {G{\"u}ltekin}, Kayhan and {Runnoe}, Jessie C. and {Koss}, Michael J.},
        title = "{AGN Triality of Triple Mergers: Detection of Faint X-Ray Point Sources}",
      journal = {\apj},
         year = 2021,
        month = feb,
       volume = {907},
       number = {2},
          eid = {71},
        pages = {71},
          doi = {10.3847/1538-4357/abce5d},
archivePrefix = {arXiv},
       eprint = {2012.00761},
 primaryClass = {astro-ph.HE},
       adsurl = {https://ui.adsabs.harvard.edu/abs/2021ApJ...907...71F}
}

@ARTICLE{2020ApJ...899..154S,
       author = {{Silverman}, John D. and {Tang}, Shenli and {Lee}, Khee-Gan and {Hartwig}, Tilman and {Goulding}, Andy and {Strauss}, Michael A. and {Schramm}, Malte and {Ding}, Xuheng and {Riffel}, Rogemar A. and {Fujimoto}, Seiji and {Hikage}, Chiaki and {Imanishi}, Masatoshi and {Iwasawa}, Kazushi and {Jahnke}, Knud and {Kayo}, Issha and {Kashikawa}, Nobunari and {Kawaguchi}, Toshihiro and {Kohno}, Kotaro and {Luo}, Wentao and {Matsuoka}, Yoshiki and {Matsuda}, Yuichi and {Nagao}, Tohru and {Oguri}, Masamune and {Ono}, Yoshiaki and {Onoue}, Masafusa and {Ouchi}, Masami and {Shimasaku}, Kazuhiro and {Suh}, Hyewon and {Suzuki}, Nao and {Taniguchi}, Yoshiaki and {Toba}, Yoshiki and {Ueda}, Yoshihiro and {Yasuda}, Naoki},
        title = "{Dual Supermassive Black Holes at Close Separation Revealed by the Hyper Suprime-Cam Subaru Strategic Program}",
      journal = {\apj},
         year = 2020,
        month = aug,
       volume = {899},
       number = {2},
          eid = {154},
        pages = {154},
          doi = {10.3847/1538-4357/aba4a3},
archivePrefix = {arXiv},
       eprint = {2007.05581},
 primaryClass = {astro-ph.GA},
       adsurl = {https://ui.adsabs.harvard.edu/abs/2020ApJ...899..154S}
}

@ARTICLE{2021ApJ...922...83T,
       author = {{Tang}, Shenli and {Silverman}, John D. and {Ding}, Xuheng and {Li}, Junyao and {Lee}, Khee-Gan and {Strauss}, Michael A. and {Goulding}, Andy and {Schramm}, Malte and {Kawinwanichakij}, Lalitwadee and {Xavier Prochaska}, J. and {Hennawi}, Joseph F. and {Imanishi}, Masatoshi and {Iwasawa}, Kazushi and {Toba}, Yoshiki and {Kayo}, Issha and {Oguri}, Masamune and {Matsuoka}, Yoshiki and {Onoue}, Masafusa and {Jahnke}, Knud and {Ichikawa}, Kohei and {Hartwig}, Tilman and {Kashikawa}, Nobunari and {Kawaguchi}, Toshihiro and {Kohno}, Kotaro and {Matsuda}, Yuichi and {Nagao}, Tohru and {Ono}, Yoshiaki and {Ouchi}, Masami and {Shimasaku}, Kazuhiro and {Suh}, Hyewon and {Suzuki}, Nao and {Taniguchi}, Yoshiaki and {Ueda}, Yoshihiro and {Yasuda}, Naoki},
        title = "{Optical Spectroscopy of Dual Quasar Candidates from the Subaru HSC-SSP program}",
      journal = {\apj},
         year = 2021,
        month = nov,
       volume = {922},
       number = {1},
          eid = {83},
        pages = {83},
          doi = {10.3847/1538-4357/ac1ff0},
archivePrefix = {arXiv},
       eprint = {2105.10163},
 primaryClass = {astro-ph.GA},
       adsurl = {https://ui.adsabs.harvard.edu/abs/2021ApJ...922...83T}
}

@ARTICLE{2013CQGra..30x4005M,
       author = {{Merritt}, David},
        title = "{Loss-cone dynamics}",
      journal = {Classical and Quantum Gravity},
         year = 2013,
        month = dec,
       volume = {30},
       number = {24},
          eid = {244005},
        pages = {244005},
          doi = {10.1088/0264-9381/30/24/244005},
archivePrefix = {arXiv},
       eprint = {1307.3268},
 primaryClass = {astro-ph.GA},
       adsurl = {https://ui.adsabs.harvard.edu/abs/2013CQGra..30x4005M}
}

@INPROCEEDINGS{2003AIPC..686..201M,
       author = {{Milosavljevi{\'c}}, Milo{\v{s}} and {Merritt}, David},
        title = "{The Final Parsec Problem}",
    booktitle = {The Astrophysics of Gravitational Wave Sources},
         year = 2003,
       editor = {{Centrella}, Joan M.},
       series = {American Institute of Physics Conference Series},
       volume = {686},
        month = oct,
    publisher = {AIP},
        pages = {201-210},
          doi = {10.1063/1.1629432},
archivePrefix = {arXiv},
       eprint = {astro-ph/0212270},
 primaryClass = {astro-ph},
       adsurl = {https://ui.adsabs.harvard.edu/abs/2003AIPC..686..201M}
}

@ARTICLE{1964PhRv..136.1224P,
       author = {{Peters}, P.~C.},
        title = "{Gravitational Radiation and the Motion of Two Point Masses}",
      journal = {Physical Review},
         year = 1964,
        month = nov,
       volume = {136},
       number = {4B},
        pages = {1224-1232},
          doi = {10.1103/PhysRev.136.B1224},
       adsurl = {https://ui.adsabs.harvard.edu/abs/1964PhRv..136.1224P}
}

@ARTICLE{2015MNRAS.449...49R,
       author = {{Rodriguez-Gomez}, Vicente and {Genel}, Shy and {Vogelsberger}, Mark and {Sijacki}, Debora and {Pillepich}, Annalisa and {Sales}, Laura V. and {Torrey}, Paul and {Snyder}, Greg and {Nelson}, Dylan and {Springel}, Volker and {Ma}, Chung-Pei and {Hernquist}, Lars},
        title = "{The merger rate of galaxies in the Illustris simulation: a comparison with observations and semi-empirical models}",
      journal = {\mnras},
         year = 2015,
        month = may,
       volume = {449},
       number = {1},
        pages = {49-64},
          doi = {10.1093/mnras/stv264},
archivePrefix = {arXiv},
       eprint = {1502.01339},
 primaryClass = {astro-ph.GA},
       adsurl = {https://ui.adsabs.harvard.edu/abs/2015MNRAS.449...49R}
}

@ARTICLE{2016Natur.532..340T,
       author = {{Thomas}, Jens and {Ma}, Chung-Pei and {McConnell}, Nicholas J. and {Greene}, Jenny E. and {Blakeslee}, John P. and {Janish}, Ryan},
        title = "{A 17-billion-solar-mass black hole in a group galaxy with a diffuse core}",
      journal = {\nat},
         year = 2016,
        month = apr,
       volume = {532},
       number = {7599},
        pages = {340-342},
          doi = {10.1038/nature17197},
archivePrefix = {arXiv},
       eprint = {1604.01400},
 primaryClass = {astro-ph.GA},
       adsurl = {https://ui.adsabs.harvard.edu/abs/2016Natur.532..340T}
}

@ARTICLE{1943ApJ....97..255C,
       author = {{Chandrasekhar}, S.},
        title = "{Dynamical Friction. I. General Considerations: the Coefficient of Dynamical Friction.}",
      journal = {\apj},
         year = 1943,
        month = mar,
       volume = {97},
        pages = {255},
          doi = {10.1086/144517},
       adsurl = {https://ui.adsabs.harvard.edu/abs/1943ApJ....97..255C}
}

@ARTICLE{2023RAA....23g5024X,
       author = {{Xu}, Heng and {Chen}, Siyuan and {Guo}, Yanjun and {Jiang}, Jinchen and {Wang}, Bojun and {Xu}, Jiangwei and {Xue}, Zihan and {Caballero}, R. Nicolas and {Yuan}, Jianping and {Xu}, Yonghua and {Wang}, Jingbo and {Hao}, Longfei and {Luo}, Jingtao and {Lee}, Kejia and {Han}, Jinlin and {Jiang}, Peng and {Shen}, Zhiqiang and {Wang}, Min and {Wang}, Na and {Xu}, Renxin and {Wu}, Xiangping and {Manchester}, Richard and {Qian}, Lei and {Guan}, Xin and {Huang}, Menglin and {Sun}, Chun and {Zhu}, Yan},
        title = "{Searching for the Nano-Hertz Stochastic Gravitational Wave Background with the Chinese Pulsar Timing Array Data Release I}",
      journal = {Research in Astronomy and Astrophysics},
         year = 2023,
        month = jul,
       volume = {23},
       number = {7},
          eid = {075024},
        pages = {075024},
          doi = {10.1088/1674-4527/acdfa5},
archivePrefix = {arXiv},
       eprint = {2306.16216},
 primaryClass = {astro-ph.HE},
       adsurl = {https://ui.adsabs.harvard.edu/abs/2023RAA....23g5024X}
}

@ARTICLE{2023A&A...678A..50E,
       author = {{EPTA Collaboration} and {InPTA Collaboration} and {Antoniadis}, J. and {Arumugam}, P. and {Arumugam}, S. and {Babak}, S. and {Bagchi}, M. and {Bak Nielsen}, A. -S. and {Bassa}, C.~G. and {Bathula}, A. and {Berthereau}, A. and {Bonetti}, M. and {Bortolas}, E. and {Brook}, P.~R. and {Burgay}, M. and {Caballero}, R.~N. and {Chalumeau}, A. and {Champion}, D.~J. and {Chanlaridis}, S. and {Chen}, S. and {Cognard}, I. and {Dandapat}, S. and {Deb}, D. and {Desai}, S. and {Desvignes}, G. and {Dhanda-Batra}, N. and {Dwivedi}, C. and {Falxa}, M. and {Ferdman}, R.~D. and {Franchini}, A. and {Gair}, J.~R. and {Goncharov}, B. and {Gopakumar}, A. and {Graikou}, E. and {Grie{\ss}meier}, J. -M. and {Guillemot}, L. and {Guo}, Y.~J. and {Gupta}, Y. and {Hisano}, S. and {Hu}, H. and {Iraci}, F. and {Izquierdo-Villalba}, D. and {Jang}, J. and {Jawor}, J. and {Janssen}, G.~H. and {Jessner}, A. and {Joshi}, B.~C. and {Kareem}, F. and {Karuppusamy}, R. and {Keane}, E.~F. and {Keith}, M.~J. and {Kharbanda}, D. and {Kikunaga}, T. and {Kolhe}, N. and {Kramer}, M. and {Krishnakumar}, M.~A. and {Lackeos}, K. and {Lee}, K.~J. and {Liu}, K. and {Liu}, Y. and {Lyne}, A.~G. and {McKee}, J.~W. and {Maan}, Y. and {Main}, R.~A. and {Mickaliger}, M.~B. and {Ni{\c{t}}u}, I.~C. and {Nobleson}, K. and {Paladi}, A.~K. and {Parthasarathy}, A. and {Perera}, B.~B.~P. and {Perrodin}, D. and {Petiteau}, A. and {Porayko}, N.~K. and {Possenti}, A. and {Prabu}, T. and {Quelquejay Leclere}, H. and {Rana}, P. and {Samajdar}, A. and {Sanidas}, S.~A. and {Sesana}, A. and {Shaifullah}, G. and {Singha}, J. and {Speri}, L. and {Spiewak}, R. and {Srivastava}, A. and {Stappers}, B.~W. and {Surnis}, M. and {Susarla}, S.~C. and {Susobhanan}, A. and {Takahashi}, K. and {Tarafdar}, P. and {Theureau}, G. and {Tiburzi}, C. and {van der Wateren}, E. and {Vecchio}, A. and {Venkatraman Krishnan}, V. and {Verbiest}, J.~P.~W. and {Wang}, J. and {Wang}, L. and {Wu}, Z.},
        title = "{The second data release from the European Pulsar Timing Array. III. Search for gravitational wave signals}",
      journal = {\aap},
         year = 2023,
        month = oct,
       volume = {678},
          eid = {A50},
        pages = {A50},
          doi = {10.1051/0004-6361/202346844},
archivePrefix = {arXiv},
       eprint = {2306.16214},
 primaryClass = {astro-ph.HE},
       adsurl = {https://ui.adsabs.harvard.edu/abs/2023A&A...678A..50E}
}

@ARTICLE{2023ApJ...951L...6R,
       author = {{Reardon}, Daniel J. and {Zic}, Andrew and {Shannon}, Ryan M. and {Hobbs}, George B. and {Bailes}, Matthew and {Di Marco}, Valentina and {Kapur}, Agastya and {Rogers}, Axl F. and {Thrane}, Eric and {Askew}, Jacob and {Bhat}, N.~D. Ramesh and {Cameron}, Andrew and {Cury{\l}o}, Ma{\l}gorzata and {Coles}, William A. and {Dai}, Shi and {Goncharov}, Boris and {Kerr}, Matthew and {Kulkarni}, Atharva and {Levin}, Yuri and {Lower}, Marcus E. and {Manchester}, Richard N. and {Mandow}, Rami and {Miles}, Matthew T. and {Nathan}, Rowina S. and {Os{\l}owski}, Stefan and {Russell}, Christopher J. and {Spiewak}, Ren{\'e}e and {Zhang}, Songbo and {Zhu}, Xing-Jiang},
        title = "{Search for an Isotropic Gravitational-wave Background with the Parkes Pulsar Timing Array}",
      journal = {\apjl},
         year = 2023,
        month = jul,
       volume = {951},
       number = {1},
          eid = {L6},
        pages = {L6},
          doi = {10.3847/2041-8213/acdd02},
archivePrefix = {arXiv},
       eprint = {2306.16215},
 primaryClass = {astro-ph.HE},
       adsurl = {https://ui.adsabs.harvard.edu/abs/2023ApJ...951L...6R}
}

@ARTICLE{2003PASP..115..763C,
       author = {{Chabrier}, Gilles},
        title = "{Galactic Stellar and Substellar Initial Mass Function}",
      journal = {\pasp},
         year = 2003,
        month = jul,
       volume = {115},
       number = {809},
        pages = {763-795},
          doi = {10.1086/376392},
archivePrefix = {arXiv},
       eprint = {astro-ph/0304382},
 primaryClass = {astro-ph},
       adsurl = {https://ui.adsabs.harvard.edu/abs/2003PASP..115..763C}
}

@ARTICLE{2020MNRAS.498..537S,
       author = {{Siwek}, Magdalena S. and {Kelley}, Luke Zoltan and {Hernquist}, Lars},
        title = "{The effect of differential accretion on the gravitational wave background and the present-day MBH binary population}",
      journal = {\mnras},
         year = 2020,
        month = oct,
       volume = {498},
       number = {1},
        pages = {537-547},
          doi = {10.1093/mnras/staa2361},
archivePrefix = {arXiv},
       eprint = {2005.09010},
 primaryClass = {astro-ph.GA},
       adsurl = {https://ui.adsabs.harvard.edu/abs/2020MNRAS.498..537S}
}

@ARTICLE{2018MNRAS.475..676S,
       author = {{Springel}, Volker and {Pakmor}, R{\"u}diger and {Pillepich}, Annalisa and {Weinberger}, Rainer and {Nelson}, Dylan and {Hernquist}, Lars and {Vogelsberger}, Mark and {Genel}, Shy and {Torrey}, Paul and {Marinacci}, Federico and {Naiman}, Jill},
        title = "{First results from the IllustrisTNG simulations: matter and galaxy clustering}",
      journal = {\mnras},
         year = 2018,
        month = mar,
       volume = {475},
       number = {1},
        pages = {676-698},
          doi = {10.1093/mnras/stx3304},
archivePrefix = {arXiv},
       eprint = {1707.03397},
 primaryClass = {astro-ph.GA},
       adsurl = {https://ui.adsabs.harvard.edu/abs/2018MNRAS.475..676S}
}

@ARTICLE{2013AJ....146..160R,
       author = {{Rusli}, S.~P. and {Erwin}, P. and {Saglia}, R.~P. and {Thomas}, J. and {Fabricius}, M. and {Bender}, R. and {Nowak}, N.},
        title = "{Depleted Galaxy Cores and Dynamical Black Hole Masses}",
      journal = {\aj},
         year = 2013,
        month = dec,
       volume = {146},
       number = {6},
          eid = {160},
        pages = {160},
          doi = {10.1088/0004-6256/146/6/160},
archivePrefix = {arXiv},
       eprint = {1310.5310},
 primaryClass = {astro-ph.CO},
       adsurl = {https://ui.adsabs.harvard.edu/abs/2013AJ....146..160R}
}

@ARTICLE{2014MNRAS.444.2700D,
       author = {{Dullo}, Bililign T. and {Graham}, Alister W.},
        title = "{Depleted cores, multicomponent fits, and structural parameter relations for luminous early-type galaxies}",
      journal = {\mnras},
         year = 2014,
        month = nov,
       volume = {444},
       number = {3},
        pages = {2700-2722},
          doi = {10.1093/mnras/stu1590},
archivePrefix = {arXiv},
       eprint = {1310.5867},
 primaryClass = {astro-ph.CO},
       adsurl = {https://ui.adsabs.harvard.edu/abs/2014MNRAS.444.2700D}
}

@ARTICLE{2006ApJ...646...49R,
       author = {{Rodriguez}, C. and {Taylor}, G.~B. and {Zavala}, R.~T. and {Peck}, A.~B. and {Pollack}, L.~K. and {Romani}, R.~W.},
        title = "{A Compact Supermassive Binary Black Hole System}",
      journal = {\apj},
         year = 2006,
        month = jul,
       volume = {646},
       number = {1},
        pages = {49-60},
          doi = {10.1086/504825},
archivePrefix = {arXiv},
       eprint = {astro-ph/0604042},
 primaryClass = {astro-ph},
       adsurl = {https://ui.adsabs.harvard.edu/abs/2006ApJ...646...49R}
}

@ARTICLE{2012ApJS..201...23E,
       author = {{Eracleous}, Michael and {Boroson}, Todd A. and {Halpern}, Jules P. and {Liu}, Jia},
        title = "{A Large Systematic Search for Close Supermassive Binary and Rapidly Recoiling Black Holes}",
      journal = {\apjs},
         year = 2012,
        month = aug,
       volume = {201},
       number = {2},
          eid = {23},
        pages = {23},
          doi = {10.1088/0067-0049/201/2/23},
archivePrefix = {arXiv},
       eprint = {1106.2952},
 primaryClass = {astro-ph.CO},
       adsurl = {https://ui.adsabs.harvard.edu/abs/2012ApJS..201...23E}
}

@ARTICLE{2015ApJS..221....7R,
       author = {{Runnoe}, Jessie C. and {Eracleous}, Michael and {Mathes}, Gavin and {Pennell}, Alison and {Boroson}, Todd and {Sigur{\dh}sson}, Steinn and {Bogdanovi{\'c}}, Tamara and {Halpern}, Jules P. and {Liu}, Jia},
        title = "{A Large Systematic Search for Close Supermassive Binary and Rapidly Recoiling Black Holes. II. Continued Spectroscopic Monitoring and Optical Flux Variability}",
      journal = {\apjs},
         year = 2015,
        month = nov,
       volume = {221},
       number = {1},
          eid = {7},
        pages = {7},
          doi = {10.1088/0067-0049/221/1/7},
archivePrefix = {arXiv},
       eprint = {1509.02575},
 primaryClass = {astro-ph.GA},
       adsurl = {https://ui.adsabs.harvard.edu/abs/2015ApJS..221....7R}
}

@ARTICLE{2017MNRAS.468.1683R,
       author = {{Runnoe}, Jessie C. and {Eracleous}, Michael and {Pennell}, Alison and {Mathes}, Gavin and {Boroson}, Todd and {Sigur{\dh}sson}, Steinn and {Bogdanovi{\'c}}, Tamara and {Halpern}, Jules P. and {Liu}, Jia and {Brown}, Stephanie},
        title = "{A large systematic search for close supermassive binary and rapidly recoiling black holes - III. Radial velocity variations}",
      journal = {\mnras},
         year = 2017,
        month = jun,
       volume = {468},
       number = {2},
        pages = {1683-1702},
          doi = {10.1093/mnras/stx452},
archivePrefix = {arXiv},
       eprint = {1702.05465},
 primaryClass = {astro-ph.GA},
       adsurl = {https://ui.adsabs.harvard.edu/abs/2017MNRAS.468.1683R}
}

@ARTICLE{2016MNRAS.463.2145C,
       author = {{Charisi}, M. and {Bartos}, I. and {Haiman}, Z. and {Price-Whelan}, A.~M. and {Graham}, M.~J. and {Bellm}, E.~C. and {Laher}, R.~R. and {M{\'a}rka}, S.},
        title = "{A population of short-period variable quasars from PTF as supermassive black hole binary candidates}",
      journal = {\mnras},
         year = 2016,
        month = dec,
       volume = {463},
       number = {2},
        pages = {2145-2171},
          doi = {10.1093/mnras/stw1838},
archivePrefix = {arXiv},
       eprint = {1604.01020},
 primaryClass = {astro-ph.GA},
       adsurl = {https://ui.adsabs.harvard.edu/abs/2016MNRAS.463.2145C}
}

@ARTICLE{2024ApJ...971L..29L,
       author = {{Liepold}, Emily R. and {Ma}, Chung-Pei},
        title = "{Big Galaxies and Big Black Holes: The Massive Ends of the Local Stellar and Black Hole Mass Functions and the Implications for Nanohertz Gravitational Waves}",
      journal = {\apjl},
         year = 2024,
        month = aug,
       volume = {971},
       number = {2},
          eid = {L29},
        pages = {L29},
          doi = {10.3847/2041-8213/ad66b8},
archivePrefix = {arXiv},
       eprint = {2407.14595},
 primaryClass = {astro-ph.GA},
       adsurl = {https://ui.adsabs.harvard.edu/abs/2024ApJ...971L..29L}
}

@ARTICLE{2017MNRAS.471.4508K,
       author = {{Kelley}, Luke Zoltan and {Blecha}, Laura and {Hernquist}, Lars and {Sesana}, Alberto and {Taylor}, Stephen R.},
        title = "{The gravitational wave background from massive black hole binaries in Illustris: spectral features and time to detection with pulsar timing arrays}",
      journal = {\mnras},
         year = 2017,
        month = nov,
       volume = {471},
       number = {4},
        pages = {4508-4526},
          doi = {10.1093/mnras/stx1638},
archivePrefix = {arXiv},
       eprint = {1702.02180},
 primaryClass = {astro-ph.HE},
       adsurl = {https://ui.adsabs.harvard.edu/abs/2017MNRAS.471.4508K}
}

@ARTICLE{2023MNRAS.522.2707S,
       author = {{Siwek}, Magdalena and {Weinberger}, Rainer and {Hernquist}, Lars},
        title = "{Orbital evolution of binaries in circumbinary discs}",
      journal = {\mnras},
         year = 2023,
        month = jun,
       volume = {522},
       number = {2},
        pages = {2707-2717},
          doi = {10.1093/mnras/stad1131},
archivePrefix = {arXiv},
       eprint = {2302.01785},
 primaryClass = {astro-ph.HE},
       adsurl = {https://ui.adsabs.harvard.edu/abs/2023MNRAS.522.2707S}
}

@ARTICLE{2023MNRAS.526.2688R,
       author = {{Rawlings}, Alexander and {Mannerkoski}, Matias and {Johansson}, Peter H. and {Naab}, Thorsten},
        title = "{Reviving stochasticity: uncertainty in SMBH binary eccentricity is unavoidable}",
      journal = {\mnras},
         year = 2023,
        month = dec,
       volume = {526},
       number = {2},
        pages = {2688-2695},
          doi = {10.1093/mnras/stad2891},
archivePrefix = {arXiv},
       eprint = {2307.08756},
 primaryClass = {astro-ph.GA},
       adsurl = {https://ui.adsabs.harvard.edu/abs/2023MNRAS.526.2688R}
}

@ARTICLE{2009ApJ...698..198G,
       author = {{G{\"u}ltekin}, Kayhan and {Richstone}, Douglas O. and {Gebhardt}, Karl and {Lauer}, Tod R. and {Tremaine}, Scott and {Aller}, M.~C. and {Bender}, Ralf and {Dressler}, Alan and {Faber}, S.~M. and {Filippenko}, Alexei V. and {Green}, Richard and {Ho}, Luis C. and {Kormendy}, John and {Magorrian}, John and {Pinkney}, Jason and {Siopis}, Christos},
        title = "{The M-{\ensuremath{\sigma}} and M-L Relations in Galactic Bulges, and Determinations of Their Intrinsic Scatter}",
      journal = {\apj},
         year = 2009,
        month = jun,
       volume = {698},
       number = {1},
        pages = {198-221},
          doi = {10.1088/0004-637X/698/1/198},
archivePrefix = {arXiv},
       eprint = {0903.4897},
 primaryClass = {astro-ph.GA},
       adsurl = {https://ui.adsabs.harvard.edu/abs/2009ApJ...698..198G}
}

@ARTICLE{2013ApJ...764..184M,
       author = {{McConnell}, Nicholas J. and {Ma}, Chung-Pei},
        title = "{Revisiting the Scaling Relations of Black Hole Masses and Host Galaxy Properties}",
      journal = {\apj},
         year = 2013,
        month = feb,
       volume = {764},
       number = {2},
          eid = {184},
        pages = {184},
          doi = {10.1088/0004-637X/764/2/184},
archivePrefix = {arXiv},
       eprint = {1211.2816},
 primaryClass = {astro-ph.CO},
       adsurl = {https://ui.adsabs.harvard.edu/abs/2013ApJ...764..184M}
}

@ARTICLE{2025arXiv250506221M,
       author = {{Mohammed}, Niana N. and {Runnoe}, Jessie C. and {Eracleous}, Michael and {Bogdanovi{\'c}}, Tamara and {Stern}, Daniel and {Simon}, Joseph and {Charisi}, Maria and {Lazio}, T. Joseph W. and {Szekerczes}, Kaitlyn and {Sigurdsson}, Steinn and {Dabbieri}, Collin},
        title = "{Testing the Hypothesis that the Quasar J0950+5128 Harbors a Supermassive Black Hole Binary}",
      journal = {arXiv e-prints},
         year = 2025,
        month = may,
          eid = {arXiv:2505.06221},
        pages = {arXiv:2505.06221},
          doi = {10.48550/arXiv.2505.06221},
archivePrefix = {arXiv},
       eprint = {2505.06221},
 primaryClass = {astro-ph.GA},
       adsurl = {https://ui.adsabs.harvard.edu/abs/2025arXiv250506221M}
}

@ARTICLE{2025arXiv250814253H,
       author = {{Holley-Bockelmann}, Kelly and {Khan}, Fazeel and {Williams}, Isaiah and {Roth}, Jaelyn and {Rizzo Smith}, Michael and {Porter}, Kaitlin and {Bellovary}, Jillian and {Derdzinski}, Andrea and {Macci{\`o}}, Andrea},
        title = "{Handy Relation Between Binary Black Hole Merger Times and Host Galaxy Properties}",
      journal = {arXiv e-prints},
         year = 2025,
        month = aug,
          eid = {arXiv:2508.14253},
        pages = {arXiv:2508.14253},
          doi = {10.48550/arXiv.2508.14253},
archivePrefix = {arXiv},
       eprint = {2508.14253},
 primaryClass = {astro-ph.GA},
       adsurl = {https://ui.adsabs.harvard.edu/abs/2025arXiv250814253H}
}

@ARTICLE{2024A&A...686A.157N,
       author = {{Nelson}, Dylan and {Pillepich}, Annalisa and {Ayromlou}, Mohammadreza and {Lee}, Wonki and {Lehle}, Katrin and {Rohr}, Eric and {Truong}, Nhut},
        title = "{Introducing the TNG-Cluster simulation: Overview and the physical properties of the gaseous intracluster medium}",
      journal = {\aap},
         year = 2024,
        month = jun,
       volume = {686},
          eid = {A157},
        pages = {A157},
          doi = {10.1051/0004-6361/202348608},
archivePrefix = {arXiv},
       eprint = {2311.06338},
 primaryClass = {astro-ph.GA},
       adsurl = {https://ui.adsabs.harvard.edu/abs/2024A&A...686A.157N}
}

@ARTICLE{1997ApJ...481..710C,
       author = {{Carollo}, C. Marcella and {Franx}, Marijn and {Illingworth}, Garth D. and {Forbes}, Duncan A.},
        title = "{Ellipticals with Kinematically Distinct Cores: V - I Color Images with WPFC2}",
      journal = {\apj},
         year = 1997,
        month = may,
       volume = {481},
       number = {2},
        pages = {710-734},
          doi = {10.1086/304060},
archivePrefix = {arXiv},
       eprint = {astro-ph/9701218},
 primaryClass = {astro-ph},
       adsurl = {https://ui.adsabs.harvard.edu/abs/1997ApJ...481..710C}
}

@ARTICLE{2013MNRAS.436.2997D,
       author = {{D'Orazio}, Daniel J. and {Haiman}, Zolt{\'a}n and {MacFadyen}, Andrew},
        title = "{Accretion into the central cavity of a circumbinary disc}",
      journal = {\mnras},
         year = 2013,
        month = dec,
       volume = {436},
       number = {4},
        pages = {2997-3020},
          doi = {10.1093/mnras/stt1787},
archivePrefix = {arXiv},
       eprint = {1210.0536},
 primaryClass = {astro-ph.GA},
       adsurl = {https://ui.adsabs.harvard.edu/abs/2013MNRAS.436.2997D}
}
\bibliographystyle{aasjournalv7}



\end{document}